\documentclass[12pt, a4paper]{article}

\usepackage{graphicx}
\usepackage{mathtext}
\usepackage{natbib}

\begin{document}

\title{Regimes of self-organized criticality in the atmospheric convection}
\author{F. Spineanu, M. Vlad, D. Palade \\
National Institute of Laser, Plasma and Radiation Physics\\
Bucharest 077125, Romania}
\maketitle

\begin{abstract}
Large scale organization in ensembles of events of atmospheric convection
can be generated by the combined effect of forcing and of the interaction
between the rising plumes and the environment. Here the \textquotedblleft
large scale\textquotedblright\ refers to the space extension that is larger
or comparable with the basic resolved cell of a numerical weather prediction
system. Under the action of external forcing like heating individual events of convection respond to the slow accumulation of vapor by a threshold-type dynamics. This is due to the a time-scale separation, between the slow drive and the fast convective response, expressed as the \textquotedblleft quasi-equilibrium\textquotedblright\ .  When there is interaction between the convection plumes, the effect is a correlated response. We show that the correlated response have many of the characteristics of the self-organized criticality (SOC). 
It is suggested that from the SOC perspective, a description of the specific dynamics induced 
by \textquotedblleft quasi-equilibrium\textquotedblright\  can be provided by models of \textquotedblleft punctuated equilibrium\textquotedblright\ . Indeed the
Bak-Sneppen model is able to reproduce (within reasonable approximation) two
of the statistical results that have been obtained in observations on the organized convection.

We also give detailed derivation of the equations connecting the probabilities of the states in the update sequence of the Bak-Sneppen model with $K=2$ random neighbors. This analytical framework allows the derivation of scaling laws for the size of avalanches, a result that guves support to the SOC interpretation of the observational data. 
\end{abstract}

\section{Introduction}

Many natural systems consist of large collections of identical sub-systems
that do not interact one with another. In many cases the ensemble (\emph{i.e.%
} the whole system) is subject to an external drive at a rate which is much
slower than the corrective reaction of any sub-system when it becomes
unstable. In some circumstances the system evolves to a type of behavior
called \textquotedblleft Self - Organized Criticality\textquotedblright\
(SOC). To introduce the basic terminology we start with a simple example: an
inflammable gas emerging with a slow rate at the flat surface of a porous
material. The burning occurs when the fraction of density of inflammable gas
in the air reaches a threshold, at the surface of the porous body. Due to
the porosity the gas will accumulate at random at the surface, eventually
creating patches of gas that emerged from neighbor pores. If above one of
such pore there is ignition, it propagates very rapidly (like an avalanche)
to all the sites within the patch, burning the gas and resetting to initial
state. Patches that are spatially separated (\emph{i.e.} non-connex) are not
affected. The dimensions of the patches (equivalently, of the avalanches of
burn) are arbitrary and can be as large as the whole surface of the porous
material. This state is similar to the criticality, without however the
system to undergo any phase transition. The state is stationary in the
statistical sense. The fluctuations (local burning events) are correlated in
space and in time with dependences on space and respectively time intervals
exhibiting long range algebraic decays. As simple as it is, this system
introduces basic elements: slow rate of feeding, threshold, fast reaction in
each column, avalanches. These are elements that describe one of the most
fundamental way of organization of systems in nature.

Each sub-system has a behavior that is characterised by a threshold. The
threshold refers to a parameter (like: density of inflammable gas, height of
a column in a sand pile, amount of water vapor in a convective column, etc.)
and separates the regime of quiet equilibrium from the active regime. The
activity consists of a fast instability (burning, toppling of sand grains
from the column, precipitation from the convection, etc.) that returns the
sub-system to the equilibrium regime, in general by removing a certain
amount of substance, of the same nature as the drive: gas, sand, water, etc.
This is transmitted to neighbors and those that are close to threshold can
themselves switch to active state, etc. This chain of influences triggered
by the instability of one initial sub-system is fast like an avalanche and
involves a set of sub-systems that were all close to threshold. It only
stops when the neighbor sub-systems are far from threshold and even if they
receive the pulse from the active ones they cannot reach the active regime.
The \textquotedblleft substance\textquotedblright\ is redistributed among
the sub-systems. This picture can easily be generalized to situations where
there is no \textquotedblleft substance\textquotedblright\ that is
transfered between the sub-systems. For example, a sub-system that switches
to active state can undergo an internal re-organization that reinstates back
the equilibrium and it simply emits a signal. Neighboring sub-systems that
receive the signal and are themselves close to threshold will switch to
active state and undergo internal re-organization, emitting signals, etc.

\bigskip

The drive originates in an external source (e.g. the radiative heating of
the land) and may affect one or several sub-systems, chosen at random.
Equally possible, the source can drive the whole ensemble of sub-systems,
uniformly or not, but always at a rate which is slow relative to the fast
reaction of any sub-system when the threshold is exceeded. Under the weak
drive the system evolves slowly to a state in which most of the sub-systems
are close to threshold. The further slow \textquotedblleft
drive\textquotedblright\ will produce avalanches of various sizes that
return the sub-systems to the equilibrium state (\emph{i.e.} under the
threshold). The system as a whole preserves this state as statistical
stationarity. This state is very similar to the state of a system that is on
the point to make a phase transition. The size of an avalanche is similar to
the length of correlation of the fluctuations. The avalanches can be
extended over all spatial scales, up to the dimension of the system and in
this analogy a correlation length which involves all the spatial extension
of the system is the signature of criticality. In the case of SOC the system
does not make a phase-transition but stops at the critical state. The SOC
state is statistically stationary so it is energetically ideal : the
activity consists of random transients (avalanches) and the system explores
the space of states specific for criticality. The system now acts under a
rule: minimum rate of entropy production. It just reacts to external
\textquotedblleft excitation\textquotedblright\ such as to keep this
statistical equilibrium.

\section{Classical view on SOC}

During the slow feeding many sub-system can reach the marginal stability,
without however becoming active. The existence of a threshold means that the
state of activity is separated from the last \textquotedblleft
bound\textquotedblright\ state, the marginally -stable one, by an interval.
A simple fluctuation, with an amplitude comparable with this gap, will
switch the sub-system from marginal stability to active state. There are
many systems that exhibit SOC. Models are of the types: (a) algorithmic,
like sand pile, Bak - Sneppen, described by a set of rules for advancing in
time the components (sub-systems); (b) analytic: like the Kardar Parisi
Zhang equation describing gradient drive $\mathbf{v}=-\mathbf{\nabla }h$ for
a scalar field $h$ that may represent the local accumulation of a quantity
(dust falling randomly on the bottom of a river). For this case the approach
is based on the Dynamic Renormalization Group. Very good references exists
on the SOC subject \citep{bak1996nature, sornette, jensen1998self}.

\section{The SOC of the atmospheric convection processes}

The physical processes that take place inside the grid cell of a finite
resolution numerical model have a fundamental role in the success of the
large scale dynamics simulation. This is the problem of parameterization,
still under active research. Part of the difficulty comes from the fact that
there is a wide variety of situations at the small scales which would have
to be represented. The diversity of physical states cannot be simply reduced
to few global charactersitics whose formal description would allow to
transfer reliable quantities to the large scale dynamics. Different physical
situations require different formalisms, with various weights placed on the
component of the small scale description: convection (either shallow or
deep), cloud distribution, entrainment mechanisms (which are dominated by
turbulence with different characteristics), detrainment, downdraft and
effects on low level convergence, etc. We should probably admit that,
instead of a single formalism for parameterization, one must consider a
variety of formalisms, each adapted to the characteristics that are dominant
at a certain state of the atmosphere in the small scale region. However this
is a heavy theoretical task. It first requires to reduce the diversity of
physical situations to a finite number, ennumerated and characterized in a
systematic way. Second, one should find a way to identify which particular
\textquotedblleft behavior\textquotedblright\ is manifested such as to
activate the adequate formalism, prepared for that particular
\textquotedblleft behavior\textquotedblright . This is a difficult
programme, but one can recognize that the stochastic parameterization seems
to partially fulfill this task, since by definition it explores a variety of
states which belong to the statistical ensemble of realizations of a field
of random physical variables. An interesting dynamics is revealed by the numerical studies based on
 \emph{cellular automata} \citep{BechtoldCellular}.

This introduction is intended to suggest that in connection with the search
for a reliable parameterization (and implicitely for the successful large
scale numerical modeling) one needs to explore the diversity of physical
situations which occur in limited areas and draw conclusions on how
different the formalism can be for the description of each of them.

\bigskip

It is not sure that this approach can be realized in practice. However it
produces interesting and possibly useful secondary results. This is because
it suggests to abandon the search for a unique formalism and encourrage to
first investigate the typologies of \textquotedblleft
behaviors\textquotedblright\ that can be identified as sufficiently
individualized and distinct. Applying this idea and looking for general
characteristics of evolutions (connecting small scales, of the order of grid
cell, and the large scales) one can identify in some circumstances
systematic elements which point to the large scale organization of
convective events. Many observational studies support this conclusion, but a
unifying concept that should be found behind this manifestation has not yet
been formulated.

Obviously, the large scale organization of convective events is not the
unique possible state of the atmosphere. Strong drive inducing large scale
response consisting of strong horizontal pressure gradients, generation of
vortical flows, jets and in general displacement of masses of air over large
distances do not fit into the characteristic state of SOC (at least in its
most common, simple, picture) and need different approaches. The
organization of the convection in the way that can be described by SOC
apperas to be just one of the possible states of the atmosphere and as such
SOC cannot claim to become the unique, paradigmatic reference of atmosphere
dynamics. But it is relevant in some particular situations and this is
supported by the fact that correlations of fluctuations of physical
quantities show algebraic decay both in space and in time. The basic
elements mentioned before as specific to SOC are a natural component of the
large scale convection systems. The \textquotedblleft
quasi-equilibrium\textquotedblright\ hypothesis recognizes the presence of
two largely separated time scales: the slow external forcing and the fast
convective response, which is similar to the corresponding property of the
sub-systems' dynamics in SOC. The mutual influence between the sites of
convection is also similar to the avalanches.

\bigskip

The important aspect of large scale organization of the convection and
precipitation has been revealed, in a more or less explicit form, in many
works. In particular, it has been accumulated evidence of a mesoscale
organization of systems of clouds. Leray and Houze (\citeyear{LearyHouze}) study
the genesis and evolution of a tropical cloud cluster. In the formation
phase the spatial characteristic consists of a line of isolated cumulonimbus
cells whose orientation is transversal on the direction of the low level
wind. Further, rain areas within individual cells merge. In this period new
convective cells are generated between and ahead of the existing cells. This
is explained by the downdrafts originating from old convection cells, which
enhance the convergence at low levels. This provides moist flux to new
cumulonimbus updrafts, enhancing their buoyancy, and producing new
convection cells. While the convection cells will eventually dissipate, new
other convections are generated. This is a propagation, in which the new
convective cells develop in front of the line of advancing precipitation
system, faster than the dissipation of the older cells at the rear. We
underline the effect that a localized convection exerts on the neighbor
sites, this being compatible with the idea that there can be propagation of
an effect in an ensemble of sites, as in an avalanche specific to SOC. The
role of mutual trigger is played by the downdraft from previous active
convection sites. The cloud cluster is characterized, at later times, by the
persistence of a large area of precipitation behind the advancing front.

From this complex picture we focus on the aspects related to the propagation
of influences between sites of convection, which bear some analogy with the
avalanche phenomena and seem to support the concept of self-organization at
criticality. We also note a particularity, that propagation in systems of
convections located at nearby sites needs a finite time interval, while in
many classical realizations of SOC the avalanche is taken as simultaneous in
all sites that are involved.

The formation of \textquotedblleft clumps\textquotedblright\ of clouds has
been discussed by Lopez (\citeyear{Lopez1977}), pointing out the random growth
before merging into large scale mesoscopic formations. We note then that the
SOC concept of avalanche, which is a correlated behavior of a set of
sub-systems that are all at, or close, to the marginal stability and switch
to active state by the effect of a influence coming from a nearby site, must
here be adapted and become the signature of the process of generation of
large scale formation of clouds out of isolated active sites of convection.
The resource of active convection are stimulated by mutual influence.

The paper of \citet*{SuBrethertonChen} on self
aggregation is a study of the generation of large clusters from isolated
convections. Various scenarios have been proposed, \emph{e.g} the
\textquotedblleft gregarious\textquotedblright\ convection \citep%
{MapesGregarious} or, the wind-induced surface heat exchange \citep%
{Emanuel1987}, \citep{NeelinHeldCook}. The large scale organization of
convection events has a multiple manifestation but at least the first phase
seems to strongly suggest SOC. The initially distinct convections interact
through the subsiding air between them and within an interval of
approximative ten days they organize into mesoscale patches of rainy air
columns. Further, the mesoscale patches of each type (rainy and respectively
dry) coalesce generating a single moist patch surrounded by dry subsiding
air. Except for the final spatial distribution the large scale propagation
of mutual influence is similar to the generation of an avalanche in a SOC
system.

A description of this propagation is offered by \citet{cruz1977}:
\textquotedblleft \emph{The radar observed progression of one hot tower is a
sequence of growing deep cumuli one ahead of the other in the direction of
the cloud motion}\textquotedblright .

\bigskip

In more precise terms, the influence consists of the change of the
environment properties in the region of the nearby site. This is done by
downdraft and precipitation.

There is a minimum time for the process of interaction to take place: the
development of the convection at the initial site, approximately half an
hour, the decay of this convection in another half of hour, the thermal
influence of the conditions for the nearby site through modification of the
vapor content and of the temperature of the air. We can take as an
elementary measure of time the interval $\delta t\sim 1\ hour$. For
comparison \citet*{plantcraig} have adopted a duration of an
individual plume $\delta t=45\ \min $. It is defined the closure time scale $%
T_{c}$, called the \emph{adjustment time} in response to a forcing. In this
interval $90\%$ of the convective available potential energy (CAPE) would be
removed if the ensemble of plumes were acting on the environment.

Now we can be interested in an estimation of the average distance between
nearby sites where convection may arise, if local conditions are favorable.
The distance between clouds, for densly packed states can be taken as the
average dimension of the cloud, assumed to be $\delta l=2\ km$. We can use
this input to estimate a speed for the propagation of the non-material
influence which consists of the fact that the convection at one site may
trigger convection at the nearby site 
\begin{equation}
v_{prop}\sim \frac{\delta l}{\delta t}=2\ km/hour  \label{eq1}
\end{equation}%
This speed is certainly different of the velocity of propagation of
gravity-inertial waves between clouds. What is important is the fact that
the SOC picture defines a new type of propagation, which is different of the
usual gravity-inertial waves generated through geostrophic adjustment.
However this propagation does not have a unique direction and does not
transport momentum or energy. Is just a trigger for convection in nearby
sites.

\bigskip

Several important works have underlined the relevance of SOC for the
atmospheric convection and have provided solid arguments. Specific to this
field, they consist of identifying algebraic decay of correlations. \citet{OlePeteRain} examines the statistics of the size of a
precipitation event, $M$. The number of events of size $M$ is%
\begin{equation}
N\left( M\right) \sim M^{-1.36}  \label{unutreizecisase}
\end{equation}%
Another observation \citep{peters2002complexity}\textbf{\ }regards the
duration between precipitation events (inter-occurrence time), which is
given in function of the duration $D$ in minutes%
\begin{equation}
N\left( D\right) \sim D^{-1.42}  \label{peters32}
\end{equation}%
We note again a close similarity with the SOC model (Bak - Sneppen) where
the exponent is $3/2$.

\bigskip

\citet*{petersneelin} propose for the intense precipitation
events a picture inspired from phase transitions, with the water vapor $w$
as the \emph{tunning parameter} (like the temperature in the magnetization)
and the precipitation rate $P\left( w\right) $ as the \emph{order parameter}%
. The slow drive is the surface heating and evaporation. The fast
dissipation of buoyancy and rain water, is the precipitating convection.
This \textquotedblleft \emph{fast dissipation by moist convection prevents
the troposphere from deviating strongly from marginal stability}%
\textquotedblright . This maintains the quasi-equilibrium which is the basic
postulate of the Arakawa Schubert parameterization \citep%
{arakawa1974interaction}.

After the critical threshold, the statistical averages have variations with
the tuning parameter $w$ as%
\begin{equation}
\left\langle P\right\rangle \left( w\right) =a\left( w-w_{c}\right) ^{\beta }
\label{eq2}
\end{equation}%
where $\beta $ is an universal exponent. After scaling with factors that are
imposed by the different climateric regions considered, the variation of the
averaged probability $\left\langle P\right\rangle $ with the difference $%
\left( w-w_{c}\right) $ shows the same slope%
\begin{equation}
\log \left\langle P\right\rangle \sim \beta \log \left( w-w_{c}\right)
\label{eq3}
\end{equation}%
with a slope $\beta \sim 0.215$.

\bigskip

\citet*{petersneelinnesbitt} analyse the mesoscale
organization of convective events from the point of view of similarity with
the statistics of percolation. The essential element that underlies the
large scale organization is a local property: the sharp increase in the rate
of precipitation beyond a certain value of the water-vapor content of the
air column. This sharp threshold and the fast reaction that follows
(precipitation) are seen as elements composing the usual scenario of a first
order phase transition, in a continuous version. Since the distinctive
component of the large scale organization in a limiting regime is the
generation of a cluster (of convective events) of the size comparable to the
system's spatial extension, it has been assumed that this is analogous to
the percolation in a two-dimensional lattice of random bonds.

Although this is a very solid argumentation in favor of a kind of phase
transition we have reserve to the proposed classification of this state as a
phase transition. Indeed there are arguments to place the large scale
organization of convective events in the same universality class as the
percolation in $2D$. However we should remember that the SOC itself, as
represented, for example, by the sand pile, belongs to the universality
class of the percolation. This is interesting in itself but does not cover
all possible interpretations that can be associated to the power law
dependence of the correlations of the fluctuating fields in a cluster of
convections. We find more appropriate to associate the generation of a
correlated cluster of convection that covers almost all the space domain
investigated, - with the divergence of the susceptibility, but, recognizing
simultaneously that correlations on all spatial scale are also possible.
Equivalently, we find an approximative state of criticality for the system.
Since however a phase transition, to a completely new phase, is not - and
cannot be seen, we identify this state as the statistical stationarity
specific to the self-organization at criticality.

\bigskip
\bigskip

\begin{figure}[h]
\centering
\includegraphics[width=0.7\linewidth]{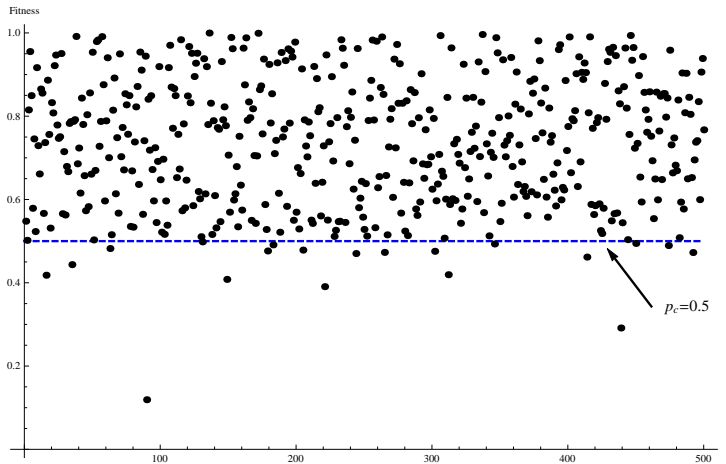}
\caption{Simulated avalanche in the Bak-Sneppen model in 1D with one random site updated; the simulation uses $n=500$ species; first true avalanche appears after $\sim 10^3 $ updates from the initial homogeneous state.}
\label{fig:1drandomneigh-avalanche}
\end{figure}

\begin{figure}[h]
\centering
\includegraphics[width=0.7\linewidth]{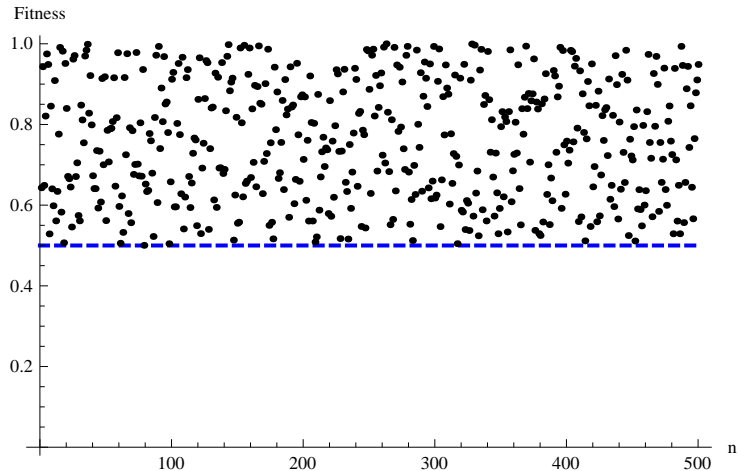}
\caption{Numerical simulation of the critical state in the Bak-Sneppen model in 1D with one random updated site; the simulation uses $n=500$ species and in the large time limit the threshold $p_c=0.5$ it is reached (blue, dashed line)}
\label{fig:1drandomneigh-treshold}
\end{figure}
\bigskip

Many factors, like the content of water in a column of air, the type of
convection (shallow or deep), the spatial correlation through the effect of
downdrafts on the environment, etc. contribute to the regime of
precipitation. Besides the complexity of the fluid and thermic processes it
is still interesting to look for a low order dynamics that would be able to
capture the essential aspects. A first suggestion comes from the fact that
the statistics of cloud clusters exhibits multiple scale organization, whose
protopype is the Continuous Time Random Walk. From the universality of
classes of statistical ensembles of random processes, the rate of occurence
of earthquakes has a similar property. The similarity between the
statistical properties of cloud sizes and of earthquake intensities has been
noted and studied previously. It is then simpler to examine the way this
statistics results from general properties of the tectonic breaking, with
phases of accumulation followed by sudden release of energy. The known
property of the earthquakes to exhibit self-organization at criticality is
therefore indirectly made plausible, \emph{i.e.} it extends to the similar
ensemble of random precipitation events. Or, there is a standard reference
for the dynamical behavior of the earthquakes in the state of SOC, the Bak -
Sneppen model.

\bigskip

\section{The SOC formulation of convection within the Bak - Sneppen model}

\subsection{Why SOC realized in a \textquotedblleft punctuated
equilibrium\textquotedblright\ system may be relevant to the organization of
the convection-precipitation events}

The model consists of an ensemble of sites (sub-systems) each characterized
by a value of a parameter which reflects the degree of adaptation
(\textquotedblleft fitness\textquotedblright ) to the environment. There is
a threshold level of fitness $\lambda $. By definition the sites whose
current level of fitness is smaller than $\lambda $ are not in equilibrium
relative to the environment and can possibly switch to an active state,
which involves the update of their fitness parameter with a new, random,
value, extracted with uniform probability from $\left( 0,1\right) $. It is
only certain that the site $k$ with the smallest fitness parameter $x_{k}$
(of course, $x_{k}<\lambda $) is subject to update and the other sites (both 
$<\lambda $ and $>\lambda $) are updated only if they are currently in a
relationship of mutual influence with the site $k$.

The approach to the state of SOC in the atmosphere may be similar to the
evolution of the \emph{gap} in the Bak Sneppen model \citep{paczuskimaslovbak}%
. The gap is the distance between zero level of fitness and the current 
\emph{average} level of fitness, which is below the threshold $\lambda $.
The gap tends to increase and the difference between the value of the
fitness at a certain time and $\lambda $ decreases. The dynamical evolution
of the gap consists of plateaux interrupted by fast events when there are
avalanches. In time the gap approaches $\lambda $ which however it will only
reach asymptotically.

\bigskip
\begin{figure}[h]
\centering
\includegraphics[width=0.7\linewidth]{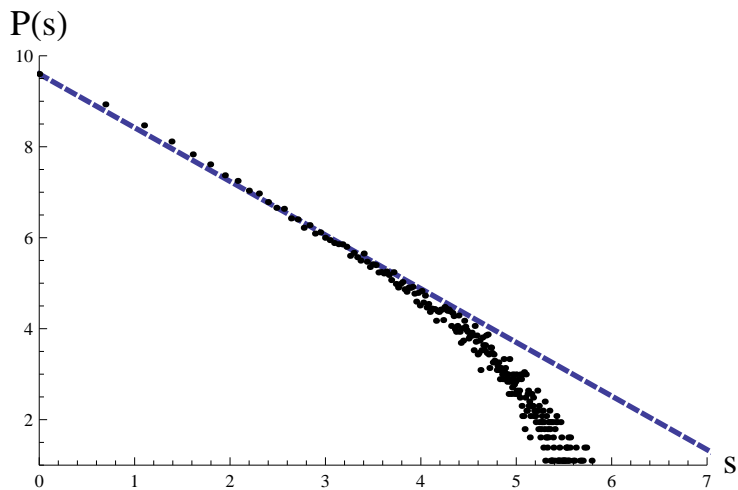}
\caption{Numerical obtained correlation for avalanche statistics $Log(P(s))$ $vs$ $Log(s)$ in the 1D Bak-Sneppen with one random updated site. The obtained exponent is $\simeq 1.32$ less than the analytical one due to finite $n$ and finite time of simulation.}
\label{fig:correlations}
\end{figure}

\begin{figure}[h]
\centering
\includegraphics[width=0.7\linewidth]{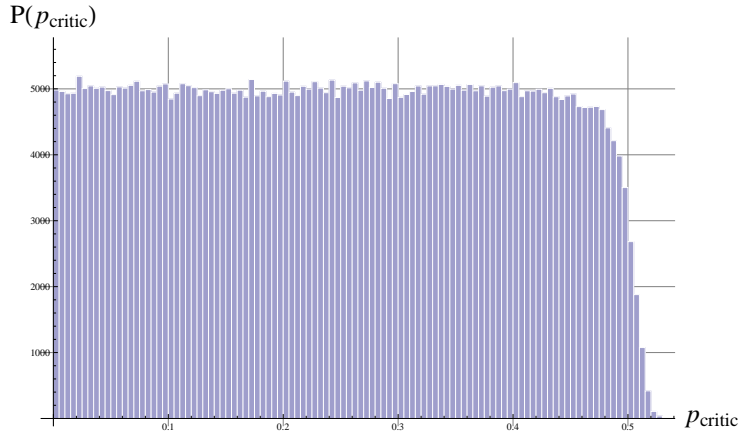}
\caption{Histogram of the minim value of fitness of every time step in the 1D Bak Sneppen with random neighbor. The Heaviside-Theta like plot has the critical value at $0.5$ as theoretical predicted. Simulation done with $n=50$ and $m=10^6$ updates}
\label{fig:histogram}
\end{figure}
\bigskip

The time evolution which precedes the SOC state therefore consists of the
increase of the average fitness parameter across the system, or, in other
words, the increase of the gap which tends to reach the $\lambda $ value. In
atmosphere this regime corresponds to the time interval when there are long
periods of inactivity interrupted by pricipitation events. The precipitation
events remove part of the water vapor from a number of sites. This
sub-ensemble of sites which are active almost simultaneously represent an
avalanche. Affecting successively various sub-ensembles of sites the process
removes part of the water vapor from almost all sites and the final result
is that the differences in the amount of water vapor decreases
progressively. The fact that the sites have similar values of this parameter
(amount of water vapor) makes possible large scale responses of the system
to a fluctuation of the drive, or, in other words, the avalanches can reach
the spatial dimension of the system. This is the signature of the
criticality. Reaching this state the system remains in a statistically
stationary state.

This is a model of \emph{punctuated equilibium. }By \emph{punctuated
equilibrium} one understands the situation where \textquotedblleft any given
small segment of the sites will experience long periods of inactivity
punctated by brief periods of violent activity\textquotedblright\ \citep%
{deboer1}. The model is shown to describe the statistics of \emph{earthquakes%
} \citep{ito}.

\bigskip

\subsection{Probability of duration of correlated convection events}

We will use the statistics of avalanches for the Bak - Sneppen model. At the
moment of update, one must identify the smallest barrier $b_{i}\in \left(
0,1\right) $ and updates the site $i$. After updating the site $i$ one
updates other $K-1$ sites, chosen at random. The model is called
\textquotedblleft random neighbors\textquotedblright\ \citep%
{paczuskimaslovbak, deboer1, deboer2}.

Our comparison is based on the following mapping. The \emph{sites} are local
atmospheric processes (attached to a small and fixed area) that are close to
produce convection. \textquotedblleft \emph{Fitness}\textquotedblright\ 
\emph{\ }is a characteristic that refers to non-activity. A site fits to the
conditions of non-activity when the conditions of convection are not
realized. The sites (local atmospheric processes) are affected by the other
sites, during their update. We now derive quantitatively few results
regarding Bak-Sneppen statistics, that seem to be close to the statistics of
the atmospheric processes.

Assume that system consists of $N$ sites, each characterized by a real
number $x_{i}\ ,\ i=1,N$ . The dynamical rule is: at each time step the $%
x_{i}$ with minimal value is replaced by a random number extracted from the
interval $\left[ 0,1\right] $ with uniform probability distribution.
Together with $x_{i}$, other $K-1$ species are replaced by random numbers,
in the same condition as $x_{i}$.

\subsection{The $K=2$ model of \emph{random neighbor}: the algorithm}

At every time step (update) it is chosen the site with the minimum barrier $%
x_{i}$ and in addition $K-1$ other sites and the fittness parameters are
updated. In this model only one other site, chosen at random, is updated 
\citep{deboer2}. The definition of an \emph{avalanche} is made in the
following steps: (1) fix a threshold barrier $\lambda $; (2) count the
number of active sites, which means count the number of sites that have a
barrier less than $\lambda $; (3) an avalanche of temporal duration $T$ is
is said to occur when there are active sites \ for consecutive $T$ temporal
steps. The duration of the avalanches has a probability with the scaling at
large $T$%
\begin{equation}
P_{aval}\left( T\right) \sim T^{-3/2}  \label{p3on2}
\end{equation}

There is a scaling low governing the time separation of the moments when the
same site will be again the minimum from all the sites. The \emph{first
return probability} $S\left( t\right) $ is defined as the probability that,
if a given site $i$ is the minimum at time $t_{0}$ it will be again minimum
- but for the first time - at time $t_{0}+t$. This means that between $t_{0}$
and $t_{0}+t$ the site $i$ has not been minimum. The probability $S\left(
t\right) $ for the random neighbor model, scales as 
\begin{equation}
S\left( t\right) \sim t^{-3/2}  \label{firstreturn}
\end{equation}

Important results on the Bak-Sneppen model can be derived analytically. We
present here and in the Appendix results that have been obtained by \citet%
{boettcherpaczuski} and \citet{deboer2}. They permit to make a comparison
with the results of \citet*{petersneelin}  in the
atmospheric convection.

We fix a real value for a parameter $\lambda $. Consider the number $n$ of
sites that have the value $x_{i}$ less than $\lambda $. Define $P_{n}\left(
t\right) $ as the probability that at time $t$ there are $n$ sites that have
value $x_{i}$ lower than $\lambda $. This probability verifies the following 
\emph{master equation}%
\begin{equation}
P_{n}\left( t+1\right) =\sum\limits_{m=0}^{N}M_{n,m}P_{m}\left( t\right)
\label{eq4}
\end{equation}%
where \citep{deboer1}%
\begin{eqnarray}
M_{n+1,n} &=&\lambda ^{2}-\lambda ^{2}\frac{n-1}{N-1}  \label{eq5} \\
M_{n,n} &=&2\lambda \left( 1-\lambda \right) +\left( 3\lambda ^{2}-2\lambda
\right) \frac{n-1}{N-1}  \nonumber \\
M_{n-1,n} &=&\left( 1-\lambda \right) ^{2}+\left( -3\lambda ^{2}+4\lambda
-1\right) \frac{n-1}{N-1}  \nonumber \\
M_{n-2,n} &=&\left( 1-\lambda \right) ^{2}\frac{n-1}{N-1}  \nonumber
\end{eqnarray}%
\begin{eqnarray}
M_{0,0} &=&\left( 1-\lambda \right) ^{2}  \label{eq6} \\
M_{1,0} &=&2\lambda \left( 1-\lambda \right)  \nonumber \\
M_{2,0} &=&\lambda ^{2}  \nonumber
\end{eqnarray}

Consider $N$ sites and fix the parameter $\lambda =1/2$. Consider that at
time $t$ there are $m$ sites for which $x_{j}<\lambda $. The probability
that at time $t$ there are $m$ such sites is $P_{m}\left( t\right) $. \ In
terms of random walk, the start of an avalanche is the start of a walker at $%
x=0$ . Later the walker stops as it reaches again $0$. This is the end of an
avalanche. One defines $P_{2n}\equiv $ probability that a walk will return
to $x=0$ at step $2n$ . To find it one starts with a more general quantity: $%
Q_{2n}=$probability that a walk started at $x=0$ returns at $x=0$
irrespective of the fact that it returns to $x=0$ several intermediate
times.This quantity is known in the theory of random walks%
\begin{equation}
Q_{2n}=\frac{1}{2^{2n}}C_{2n}^{n}  \label{eq7}
\end{equation}%
Since $Q_{2n}$ is a collection of $P_{2n}$, one can use the generating
functions%
\begin{equation}
1+\sum\limits_{n=1}^{\infty }Q_{2n}z^{2n}=\frac{1}{1-\sum\limits_{n=1}^{%
\infty }P_{2n}z^{2n}}  \label{eq8}
\end{equation}%
The left hand side is explicit and gives%
\begin{equation}
1-\sum\limits_{n=1}^{\infty }P_{2n}z^{2n}=\sqrt{1-z^{2}}  \label{eq10}
\end{equation}%
from where%
\begin{eqnarray}
P_{2n} &=&\frac{\left( 2n-3\right) !!}{\left( 2n\right) !!}\approx \sqrt{%
\frac{2}{\pi }}\frac{1}{\left( 2n\right) ^{3/2}}\ \left( \textrm{large }%
n\right)  \label{eq11} \\
&\sim &\tau ^{-3/2}  \nonumber
\end{eqnarray}%
This is the scaling of the duration of an avalanche. This is not far from
the result of Ole Peters Eq.(\ref{unutreizecisase}) in the statistics of the
atmospheric precipitation.

\bigskip

\section{The Gierer - Meinhardt model of clusters and spikes in clusters
(slow activator fast inhibitor). \emph{Spotty - spiky} solutions}

We now discuss briefly a continuous model that has some similarity to the
Bak-Sneppen model. In the realization of the algorithm of Bak-Sneppen SOC
for the convection - precipitation case we introduce: (1) the function A : 
\emph{activity at the location (x,y)}; (2) the function H: degree of chances
for instability, or: \emph{potential instability}, which is the water vapor
content. In any point, the \textquotedblleft barrier\textquotedblright\
against an update is determined as the inverse of the content of water
vapors, =1/H. These two functions A and H have a mutual relationship which
is similar to the activator and inhibitor in the Gierer Meinhardt (GM) model 
\citep{Meinhardt}. The Bak-Sneppen model is discrete and algorthmic, the
Gierer Meinhardt model is space and time continuous. There are elements that
suggests the adoption of the GM model as a low-order, continuous limit of
the ensemble of precipitation events. The GM model have spotty - spiky
solutions which can be associated to extreme events.

\bigskip

The current approach to the to local convection dynamics is originated in the mass-flux parameterization
with different developments \citep{YanoDAOsubmitted}.
Assume there are centers of convection in the surface of a grid box. There
is a parameter (called \emph{barrier}) representing the degree of fitness of
the local sub-system to the environment. For every point (sub - system) the
value of that parameter is a random number. We examine the possible
similarity by starting from the Bak-Sneppen dynamics. (1) The sub-system
with the lowest value of the parameter (barrier) is found. This may be the
point with the highest degree of Column Water Vapor content, since this
means high chance for instability. A simplified representation is to
consider the \emph{inverse} of the amount of Column Water Vapor looks
similar to 
\begin{equation}
\textrm{barrier}\sim \frac{1}{CWV}\sim \frac{1}{H}  \label{eq21}
\end{equation}%
which appears on the Gierer Mainhardt model. The \emph{smallest barrier}
means highest chance to become unstable and then to be \emph{updated}. The
nonlinearity is the GM model is%
\begin{equation}
\frac{A^{2}}{H}\sim \left( \textrm{barrier}\right) \times A^{2}  \label{eq22}
\end{equation}%
(2) The sub-system is mutated into a different state. A new barrier is
atributted to this new sub-system. The new value of the barrier is taken at
random from the same set $\left( 0,1\right) $ with uniform probability. This
is equivalent to assume that some precipitation has been produced and after
that the Column Water Vapor is different in that specific sub-system. (3)
Other sub-systems are affected by this mutation. They are also updated with
the same occasion. The update is random, with \emph{new barriers} from the
set $\left( 0,1\right) $. This may be interpreted as follows: \ other
centers of convection has been influenced by the downdraft from the main
sub-system. (4) The update of the connected species is \emph{annealed}: the
species that are chosen to be updated are chosen \emph{anew} every time it
changes. We can imagine that other centers of convection are affected by the
update of the center that was active at the current moment.

The equations%
\begin{eqnarray}
\frac{\partial A}{\partial t} &=&\varepsilon ^{2}\Delta A-A+\frac{A^{2}}{H}
\label{eq24} \\
\tau \frac{\partial H}{\partial t} &=&D\Delta H-H+A^{2}  \nonumber
\end{eqnarray}%
where $A,H>0$, with%
\begin{eqnarray}
A &\equiv &\textrm{concentration of activator at point }\left( x,y,t\right)
\label{eq26} \\
H &\equiv &\textrm{concentration of inhibitor at point }\left( x,y,t\right) 
\nonumber
\end{eqnarray}

The solutions show function \emph{activator} $A$ concentrated in $K$ points,
in different location of the domain $\Omega $. There is a phenomenon of 
\emph{solution concentration} for $\varepsilon \rightarrow 0$ the peaks
become more and more narrow and at the limit they are the points themselves.

Looking for the comparison with the Bak - Sneppen dynamics, we have%
\begin{eqnarray}
\frac{\partial A}{\partial t} &=&-A\left( 1-\frac{A}{H}\right)  \label{eq27}
\\
A &\sim &\exp \left[ -\left( 1-\frac{A}{H}\right) t\right]  \nonumber
\end{eqnarray}%
and we comment as follows%
\begin{equation}
A\sim \exp \left[ -t\right]  \label{eq28}
\end{equation}%
if there would be no other term, \emph{i.e.} no factor $\left( 1-\frac{A}{H}%
\right) $. Then the time constant is%
\begin{equation}
A\sim \exp \left[ -\frac{t}{\tau }\right] \ \ \textrm{with}\ \ \tau =1
\label{eq29}
\end{equation}%
However we have%
\begin{eqnarray}
A &\sim &\exp \left[ -\left( 1-\frac{A}{H}\right) t\right]  \label{eq30} \\
A &\sim &\exp \left[ -\frac{t}{\tau _{H}}\right]  \nonumber \\
\textrm{where}\ \ \tau _{H} &=&\frac{1}{1-\frac{A}{H}}>\tau  \nonumber
\end{eqnarray}%
so the time of decay of the \emph{activator} is longer when the nonlinear
term is included. Assume for the moment that%
\begin{equation}
A/H<1  \label{eq31}
\end{equation}%
If the $H$ function is very large, which means high content of vapor (large
CWV) then $1-\frac{A}{H}\rightarrow 1$ and $\tau _{H}\searrow \tau $ and the
decay of the local value of the \emph{activator} $A$ is again fast. The
large amount of vapor induces instability and favors the local reduction of
the \emph{activator} function.

Similarly we can evaluate qualitatively the role of the current value of the 
\emph{activator}. If initially there is a high $A$ (but still $A/H<1$) then
the time of decay $\tau _{H}$ is longer than $\tau $. The continuous decay
of $A$ makes $1-A/H$ to approach $1$ and the decay becomes faster, $\tau
_{H}\searrow \tau $. This underlines the nonlinear effect of $A$: High value
of the activator are more persistent. The decay of the \emph{activator} $A$
begins by having a slow rate but the rate accelerates in time.

Within the range where%
\begin{equation}
A/H<1  \label{eq32}
\end{equation}%
the roles of large $A$ and $H$ are opposite: large initial $A$ tends to
create persistence of $A$, but large $H$ induces instability and faster
decay of $A$.

Now, let us consider the range%
\begin{equation}
\frac{A}{H}>1  \label{eq33}
\end{equation}

This range can be reached by decreasing $H$ which means reducing the amount
of vapor content. Then the \emph{activator} $A$ increases exponentially.
When $A$ grows and enters the regime $A/H>1$ the growth is self-accelerating
since the coefficient of the exponent grows as $A$ itself.

Or, the regime can be obtained for higher $A$. The same self-acceleration
occurs.

This leads to high concentration of the \emph{activator }$A$ in few
localised regions, where it happened that $A>H$ and growth with
self-amplification has occured.

In the rest, the quantity of \emph{activator} $A$ is smaller than the vapor
content $H$ and there is quiet state, with no dynamics. $H$ may be seen as a 
\emph{passive inhibitor}, in the sense that its presence means that there
are chances for instability but the instability has not yet started.

\bigskip

If a mapping to the atmospheric convection is plausible we will name (a) the
function $A$ is \emph{activity at the location }$\left( x,y\right) $; and
(2) the function $H$ is degree of chances for instability: \emph{potential
instability}. When the degree of chance of instability (\emph{potential
instability}) is larger than the local activity $\left( H>A\right) $ the
activity has not yet started. The activity $A$ is low. The so-called \emph{%
inhibitor} $H$ does NOT suppress the activity but is just a measure in which
this one has not yet started.

When the \emph{potential instability} is lower than the \emph{activity}, $%
H<A $ this means that the activity has started in that point and the growth
of the activity is very efficient.

\bigskip

\begin{figure}[ht]
\centering
\includegraphics[width=0.7\linewidth]{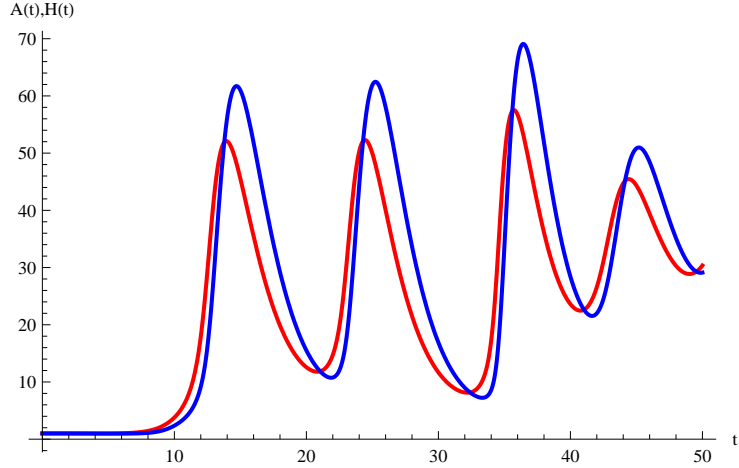}
\caption{Time variation of the two functions $A(x,y)$ (activity)
 and $H(x,y)$ (column vapor water content) resulting from the $1D$ Gierer-Meinhardt
model. Oscillatory regime}
\label{oscillatory}
\end{figure}

\begin{figure}[ht]
\centering
\includegraphics[width=0.7\linewidth]{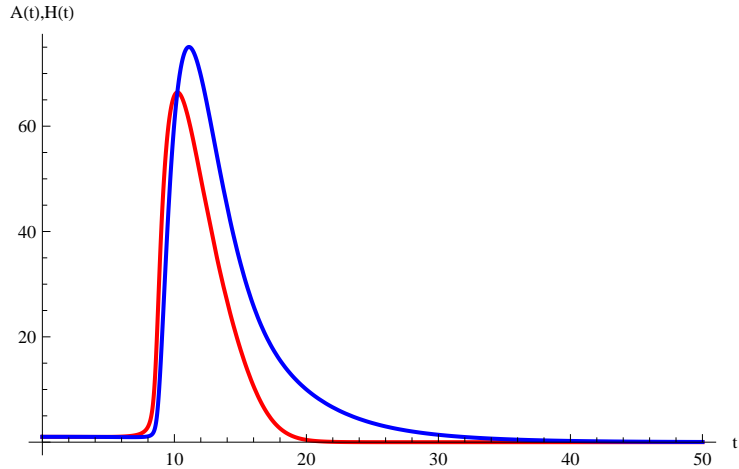}
\caption{Same as Figure (\ref{oscillatory}). A regime of fast compensation and decay.}
\label{compensation}
\end{figure}

\begin{figure}[ht]
\centering
\includegraphics[width=0.7\linewidth]{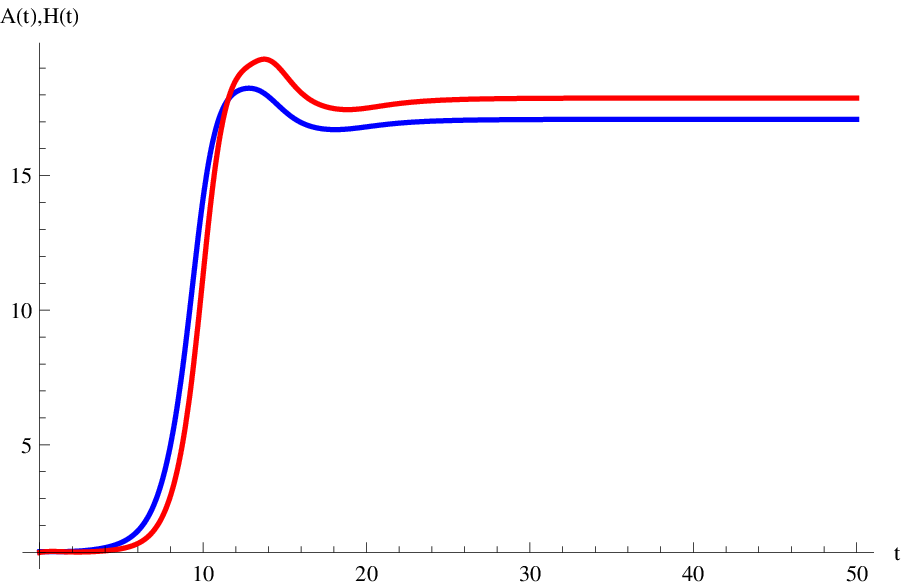}
\caption{Same as Figure (\ref{oscillatory}). A regime of saturation.}
\label{saturation}
\end{figure}
\bigskip

Now, what is the connection between this system and the Bak - Sneppen model?

The Bak - Sneppen dynamics rises continuously the level of \emph{fitness},
the value of the parameter that shows the degree of \emph{adaptation} to the
environment. This means that the points that have been updated are now more
stable than they were immediately after the random initialization.

This however does not mean stability. It only means that the degree of
stability (or fitness) has become rather similar for all the sub-systems of
the system. There are no more large discrepancies that makes of the update a
localised and isolated dynamics. Now a change can affect a large number of
sub-systems and creates an \emph{avalanche}. The system is in a critical
state and it has reached that state by its own dynamics (is self - organized
at criticality).

The degree of fitness is the \textquotedblleft barrier\textquotedblright\ of
the sub-system against mutation. It is then the inverse of the amount of
vapor, 
\begin{eqnarray}
&&\textrm{"barrier" against mutation}  \label{eq35} \\
&=&\textrm{fitness\ }=\frac{1}{CWV}=\frac{1}{H}  \nonumber
\end{eqnarray}%
is the inverse of the \emph{inhibitor}. If there is large CWV (vapor) $H$ in
a point then this means that there is low fitness, low \textquotedblleft
barrier\textquotedblright\ against mutation and the chances of instability
are large. However when the vapor $H$ is measured there is \emph{no} or low
activity $A$ in that point. The operation of \emph{update} consists of
conversion of some vapor $H$ into \emph{activity} $A$: the vapor $H$
decreases and the activity $A$ increases, the barrier is increased as
required by the Bak - Sneppen dynamics. The activity $A$ may increase but
not sufficiently such that $A>H$. Then what we get is just another landscape
of $A\left( x,y\right) $ and $H\left( x,y\right) $ but not spiky solution.

After the update we have a smaller $H$ (less vapors) in the updated
locations, which means higher \emph{barriers}. The activity has also
increased to a certain extent in those points.

The standard Bak - Sneppen random dynamics has been realised by a random
factor of conversion from the vapor content $H$ to the activity $A$.

After a sequence of such updates the activity $A$ will be higher than
initially, and with all points being at comparable degrees of activity. In
the same time, the vapor has decreased in almost all points (\emph{i.e.} the
barriers have increased for those points).

\subsection{Spikes in the solutions of the Gierer - Mainhardt model}

The system GM in $2D$ , for $t>0$ and boundary conditions%
\begin{equation}
\partial _{n}A=0\ \ ,\ \ \partial _{n}H=0\ \ \textrm{for}\ x\in \partial \Omega
\label{eq41}
\end{equation}%
exhibits spike solution that will tend to $\infty $ for $\varepsilon
\rightarrow 0$. We have solved these equations and indeed we have found that
there is concentration of the activity in localizes areas where it can have
large amplitudes. This hsould require a more detailed understanding of the

\section{Conclusions}

The atmospheric convection has many manifestations and a unique description of the specific regimes is difficult. However there are situations where
the atmospheric convection can be seen as a part of the complex phenomenon with the essential characteristics of the self-organized criticality: the slow driving, the fast reaction when there is departure from quasi-equilibrium, the interaction between neighbor sites and the formation of large-scale ensembles with correlated response. 
The correlations of fluctuations of SOC systems should exhibit universal scaling in space and time and this is indeed found for the SOC regimes of convection. Observational data are compatible with the results of analytical derivation based on a paradigmatic example of \emph{punctuated equilibrium} system, the Bak-Sneppen model. We provide detail analytical derivation of the equations for the probabilities of states of the sequential update of the Bak-Sneppen model with $k=2$ random neighbors.

We present a possible extension to continuum dynamics, by the model Gierer-Meinhardt. The numerical investigation confirms, to a certain extent, the expectation: the system evolves by a continuous mutual control of an \emph{inhibitor} and an \emph{activator} variables. It also presents spiky solutions that may be of interest in the investigation of the extreme dynamics. 

In conclusion, SOC appears to be a possible ground for construction of a coherent perspective on the large scale organization of convection.

\bigskip

\textbf{Acknowledgments}. This work has been partly supported by the grant
ERC-Like nr.4/2012 of UEFISCDI Romania. The exchange of ideas in the COST
ES0905 collaboration is particularly acknowledged.

\bigskip

\section{Appendix.Calculation of the probabilities of the states for
Bak-Sneppen dynamics at $K=2$ random neighbor}

\subsection{Definition and calculation of the transition probabilities}

Exact analytic results exist for the time-dependent statistical
characteristics of the Bak - Sneppen model. These have been obtained by \citet%
{deboer1}, \citet{deboer2}, \citet{paczuskimaslovbak}. In this Appendix we
start from these original works and provide details of the calculations
leading to the system of equation connecting the probabilities of the states
before and after update.

The case that will be examined has $K=2$. This means that besides the sites
with the smallest value $x_{i}$ only just another one $x_{l}$ is changed by
replacing $x_{l}$ with a new $x_{l}^{\prime }$, random, value. The site $%
x_{l}$ is chosen \emph{at random}.

We fix a real value for the parameter $\lambda $. Consider the number $n$ of
sites that have the value $x_{i}$ less than $\lambda $. Define%
\begin{equation}
P_{n}\left( t\right)   \label{a1}
\end{equation}%
as the probability that at time $t$ there are $n$ sites that have value $%
x_{i}$ lower than $\lambda $.

This probability verifies the following \emph{master equation}%
\begin{equation}
P_{n}\left( t+1\right) =\sum\limits_{m=0}^{N}M_{n,m}P_{m}\left( t\right) 
\label{a2}
\end{equation}%
where%
\begin{eqnarray}
M_{n+1,n} &=&\lambda ^{2}-\lambda ^{2}\frac{n-1}{N-1}  \label{a3} \\
M_{n,n} &=&2\lambda \left( 1-\lambda \right) +\left( 3\lambda ^{2}-2\lambda
\right) \frac{n-1}{N-1}  \nonumber \\
M_{n-1,n} &=&\left( 1-\lambda \right) ^{2}+\left( -3\lambda ^{2}+4\lambda
-1\right) \frac{n-1}{N-1}  \nonumber \\
M_{n-2,n} &=&\left( 1-\lambda \right) ^{2}\frac{n-1}{N-1}  \nonumber
\end{eqnarray}%
\begin{eqnarray}
M_{0,0} &=&\left( 1-\lambda \right) ^{2}  \label{a4} \\
M_{1,0} &=&2\lambda \left( 1-\lambda \right)   \nonumber \\
M_{2,0} &=&\lambda ^{2}  \nonumber
\end{eqnarray}

It is assumed that the total number of sites in the system is $N$. At time $%
t $ there are $m$ sites for which $x_{j}<\lambda $. The probability that at
time $t$ there are $m$ such sites is $P_{m}\left( t\right) $.

\bigskip

Now we apply the algorithmic change at step $t$.

Identify the site $k$ with the lowest value $x_{k}$ from the set of $m$
sites. This is made with probability $1/m$ because any of the $m$ sites can
be at this state.%
\begin{equation}
\textrm{probability that the site }k\textrm{ has }x_{k}\textrm{ lower than }%
\lambda \textrm{ is\ }\frac{1}{m}  \label{a5}
\end{equation}

Replace $x_{k}$ by the result $x_{k}^{\prime }$ of extracting a random
number from $\left[ 0,1\right] $ with uniform probability%
\begin{equation}
\textrm{site }k\ :x_{k}\rightarrow x_{k}^{\prime }  \label{a6}
\end{equation}%
Now we examine the two possibilities.

\bigskip

A. There is a chance that the new $x_{k}^{\prime }$ value is larger than $%
\lambda $%
\begin{eqnarray}
&&\textrm{probability that the site }k\textrm{ with minimum }x_{k}\textrm{
(principal) is updated}  \label{a7} \\
&&\textrm{to a value }x_{k}^{\prime }\textrm{ which is greater than }\lambda  
\nonumber \\
&=&\textrm{ }1-\lambda   \nonumber
\end{eqnarray}%
since the distribution is uniform on $\left[ 0,1\right] $ and the
probability is given by the length of the interval. In this case the number
of sites with value $x$ smaller than $\lambda $ at the next time step $t+1$
is smaller by one unit%
\begin{equation}
m\rightarrow m-1  \label{a8}
\end{equation}

Simultaneously we have to identify a site $l$ that is in interaction with
the site $k$. This is done under the assumption that this other site is any
of the available $N-1$ sites remaining, at random, with uniform probability.
Therefore we would be tempted to say that the probability of choosing the
site $l$ from the rest of $N-1$ sites is $1/\left( N-1\right) $.

However we must distinguish between the possibilities that $l$ has its $%
x_{l} $ smaller or larger than $\lambda $. Then we divide this step into two
branches:

\bigskip

A.1 Assume that 
\begin{equation}
\textrm{secondary }x_{l}<\lambda   \label{a9}
\end{equation}%
Then it is one of the $m$ sites that are all characterized at $t$ as having $%
x_{i}<\lambda $. (To this family belongs also the main site, the lowest, $%
x_{k}$).

The probability for this is%
\begin{eqnarray}
&&\frac{m-1}{N-1}\left( \textrm{probability to choose }l\textrm{ between the }m-1%
\textrm{ sites with }x_{i}<l\right)   \label{a10} \\
&&m-1\textrm{ sites are remaining after the site }x_{k}\textrm{ is special, the
lowest}  \nonumber
\end{eqnarray}%
Then we have two possible evolutions:

\bigskip

A.1.1 The update of the site $l$ is such that the new value of $x_{l}$ is
larger than $\lambda $.%
\begin{equation}
x_{l}\rightarrow x_{l}^{\prime }>\lambda   \label{a11}
\end{equation}%
with%
\begin{eqnarray}
&&\textrm{probability that the secondary site }l\textrm{ belongs to the set }m-1%
\textrm{ and}  \label{a12} \\
&&\textrm{after update }x_{l}^{\prime }\textrm{ is grater than }\lambda \textrm{ is%
}  \nonumber \\
&&\frac{m-1}{N-1}\times \left( 1-\lambda \right)   \nonumber
\end{eqnarray}%
This update of the other (interacting) site reduces the number of sites in
the initial set $m$ with one unit. This reduction comes after the first
reduction with one unit, made by the the principal site, $x_{k}$, the lowest
at $t$, which we assumed has been updated to $x_{k}^{\prime }>\lambda $.

Assume that we have $m$ sites that are initially under $\lambda $. Due to
these two updates, the number $m$ of sites having $x<\lambda $ changes as 
\begin{eqnarray}
x_{k} &\rightarrow &x_{k}^{\prime }\ :\ m\rightarrow m-1\ \ \left( \textrm{the
main site escapes to }>\lambda \right)   \label{a13} \\
x_{l} &\rightarrow &x_{l}^{\prime }\ :\ m-1\rightarrow m-2\ \ \left( \textrm{%
the secondary site escapes to }>\lambda \right)   \nonumber
\end{eqnarray}%
the number of sites with $x^{\prime }s$ less than $\lambda $ is%
\begin{eqnarray}
&&\textrm{number of sites with }x\textrm{ less than }\lambda \textrm{ , at time }%
t+1\textrm{ is}\   \label{a14} \\
\left( m-1\right) -1 &=&m-2  \nonumber
\end{eqnarray}%
The transition is 
\begin{equation}
P_{m}\left( t\right) \rightarrow P_{m-2}\left( t+1\right)   \label{a15}
\end{equation}%
which means a contribution to%
\begin{equation}
P_{m-2}\left( t+1\right)   \label{a16}
\end{equation}%
and the contribution is%
\begin{eqnarray}
&&\left( 1-\lambda \right) \times \frac{m-1}{N-1}\times \left( 1-\lambda
\right)   \label{a17} \\
&=&\frac{m-1}{N-1}\left( 1-\lambda \right) ^{2}  \nonumber
\end{eqnarray}

\bigskip

A.1.2. The update of the secondary (interacting) site $l$ is such that the
new value of $x_{l}$ is smaller than $\lambda $.%
\begin{equation}
x_{l}\rightarrow x_{l}^{\prime }<\lambda   \label{a18}
\end{equation}%
with%
\begin{eqnarray}
&&\textrm{probability that the secondary site }l\textrm{ belongs to the set }m-1%
\textrm{ and}  \label{a19} \\
&&\textrm{after update }x_{l}^{\prime }\textrm{ is less than }\lambda \textrm{ is}
\nonumber \\
&&\frac{m-1}{N-1}\times \lambda   \nonumber
\end{eqnarray}%
which maintains the number of sites in the initial set $m$ . Due to these
two updates 
\begin{eqnarray}
x_{k} &\rightarrow &x_{k}^{\prime }\ :\ m\rightarrow m-1\ \ \left( \textrm{the
lowest site escapes to }>\lambda \right)   \label{a20} \\
x_{l} &\rightarrow &x_{l}^{\prime }\ :\ m-1\rightarrow m-1\ \ \left( 
\begin{array}{c}
\textrm{the secondary site is not from the }m\textrm{ family} \\ 
\textrm{and remains ouside this family}%
\end{array}%
\right)   \nonumber
\end{eqnarray}%
the number of sites with $x^{\prime }s$ less than $\lambda $ is%
\begin{eqnarray}
&&\textrm{number of sites with }x\textrm{ less than }\lambda \textrm{ , at time }%
t+1\textrm{ is}\   \label{a21} \\
&&m-1  \nonumber
\end{eqnarray}%
The transition represented by these updates is%
\begin{equation}
P_{m}\left( t\right) \rightarrow P_{m-1}\left( t+1\right)   \label{a22}
\end{equation}%
which means a contribution to%
\begin{equation}
P_{m-1}\left( t+1\right)   \label{a23}
\end{equation}%
and the contribution is%
\begin{eqnarray}
&&\left( 1-\lambda \right) \times \frac{m-1}{N-1}\times \lambda   \label{a24}
\\
&=&\frac{m-1}{N-1}\left( 1-\lambda \right) \lambda   \nonumber
\end{eqnarray}

\bigskip

A.2. Now assume that the secondary, interacting, site $x_{l}$ has at $t$ a
barrier which is higher than the limit $\lambda $, 
\begin{equation}
\textrm{secondary }x_{l}>\lambda   \label{a25}
\end{equation}%
Then it is one of the $N-m$ sites that are all characterized at $t$ as
having $x_{i}>\lambda $.

The probability for this is%
\begin{equation}
\frac{N-m}{N-1}\left( \textrm{probability to choose }l\textrm{ between the }N-m%
\textrm{ sites with }x_{i}>l\textrm{ at }t\right)   \label{a26}
\end{equation}%
Then we have two possible evolutions:

\bigskip

A.2.1 The update of the (interacting) site $l$ (which belonged to $N-m$) is
such that the new value of $x_{l}$ is larger than $\lambda $.%
\begin{equation}
x_{l}\rightarrow x_{l}^{\prime }>\lambda   \label{a27}
\end{equation}%
with%
\begin{eqnarray}
&&\textrm{probability that the secondary site }l\textrm{ belongs to the set }N-m%
\textrm{ and}  \label{a28} \\
&&\textrm{after update }x_{l}^{\prime }\textrm{ is grater than }\lambda \textrm{ is%
}  \nonumber \\
&&\frac{N-m}{N-1}\times \left( 1-\lambda \right)   \nonumber
\end{eqnarray}%
which keeps constant the number of sites in the initial set $m$. Due to
these two updates 
\begin{eqnarray}
x_{k} &\rightarrow &x_{k}^{\prime }\ :\ m\rightarrow m-1\ \left( \textrm{the
lowest site escapes to higher }\lambda \right)   \label{a29} \\
x_{l} &\rightarrow &x_{l}^{\prime }\ :\ m-1\rightarrow m-1\ \left( 
\begin{array}{c}
\textrm{the secondary site does not belong to }m-1 \\ 
\textrm{and it does not enter under }\lambda 
\end{array}%
\right)   \nonumber
\end{eqnarray}%
the number of sites with $x^{\prime }s$ less than $\lambda $ is%
\begin{eqnarray}
&&\textrm{number of sites with }x\textrm{ less than }\lambda \textrm{ , at time }%
t+1\textrm{ is}\   \label{a30} \\
&&m-1  \nonumber
\end{eqnarray}%
which means a transition 
\begin{equation}
P_{m}\left( t\right) \rightarrow P_{m-1}\left( t+1\right)   \label{a31}
\end{equation}%
through a contribution to%
\begin{equation}
P_{m-1}\left( t+1\right)   \label{a32}
\end{equation}%
and the contribution is%
\begin{eqnarray}
&&\left( 1-\lambda \right) \times \frac{N-m}{N-1}\times \left( 1-\lambda
\right)   \label{a33} \\
&=&\frac{N-m}{N-1}\left( 1-\lambda \right) ^{2}  \nonumber
\end{eqnarray}

\bigskip

A.2.2. The update of the (interacting) site $l$ (which belonged to $N-m$,
having $>\lambda $) is such that the new value of $x_{l}$ is smaller than $%
\lambda $.%
\begin{equation}
x_{l}\rightarrow x_{l}^{\prime }<\lambda   \label{a35}
\end{equation}%
with%
\begin{eqnarray}
&&\textrm{probability that the secondary site }l\textrm{ belongs to the set }N-m%
\textrm{ and}  \nonumber \\
&&\textrm{after update }x_{l}^{\prime }\textrm{ is less than }\lambda \textrm{ is}
\nonumber \\
&&\frac{N-m}{N-1}\times \lambda   \label{a36}
\end{eqnarray}%
which increases the number of sites in the initial set $m$ with one unit.
Due to these two updates 
\begin{eqnarray}
x_{k} &\rightarrow &x_{k}^{\prime }\ :\ m\rightarrow m-1\ \left( \textrm{the
main site escapes to }>\lambda \right)   \label{a37} \\
x_{l} &\rightarrow &x_{l}^{\prime }\ :\ m-1\rightarrow m-1+1\ \left( 
\begin{array}{c}
\textrm{the secondary site did not belong to }m\textrm{ family} \\ 
\textrm{but now comes under }\lambda 
\end{array}%
\right)   \nonumber
\end{eqnarray}%
the number of sites with $x^{\prime }s$ less than $\lambda $ is%
\begin{eqnarray}
&&\textrm{number of sites with }x\textrm{ less than }\lambda \textrm{ , at time }%
t+1\textrm{ is}\   \label{a38} \\
&&m  \nonumber
\end{eqnarray}%
[one leaves ($x_{k}$) and other comes ($x_{l}$) ].

This case connects the situations%
\begin{equation}
P_{m}\left( t\right) \rightarrow P_{m}\left( t+1\right)   \label{a39}
\end{equation}%
which means a contribution to%
\begin{equation}
P_{m}\left( t+1\right)   \label{a40}
\end{equation}%
and the contribution is%
\begin{eqnarray}
&&\left( 1-\lambda \right) \times \frac{N-m}{N-1}\times \lambda   \label{a41}
\\
&=&\frac{N-m}{N-1}\left( 1-\lambda \right) \lambda   \nonumber
\end{eqnarray}

\bigskip

B. There is a chance that the new $x_{k}^{\prime }$ value is smaller than $%
\lambda $%
\begin{eqnarray}
&&\textrm{probability that the site }k\textrm{ with minimum }x_{k}\textrm{
(principal) is updated}  \nonumber \\
&&\textrm{to a value }x_{k}^{\prime }\textrm{ which is smaller than }\lambda  
\nonumber \\
&=&\textrm{ }\lambda   \label{a42}
\end{eqnarray}%
In this case the number of sites with value $x$ smaller than $\lambda $ at
the next time step $t+1$ does not change%
\[
m\rightarrow m
\]

Simultaneously we have to identify a site $l$ that is in interaction with
the site $k$. This is done under the assumption that this other site is any
of the available $N-1$ sites remaining. Therefore we would be tempted to say
that the probability of choosing the site $l$ from the rest of $N-1$ sites
is $1/\left( N-1\right) $.

However we must distinguish between the possibilities that $l$ has its $%
x_{l} $ smaller or larger than $\lambda $. Then we divide this step into two
branches:

\bigskip

B.1 Assume that 
\begin{equation}
\textrm{secondary (interacting) site }x_{l}<\lambda   \label{a43}
\end{equation}%
Then it is one of the $m$ sites that are all characterized at $t$ as having $%
x_{i}<\lambda $.

The probability for this is%
\begin{equation}
\frac{m-1}{N-1}\left( \textrm{probability to choose }l\textrm{ between the }m%
\textrm{ sites with }x_{i}<l\right)   \label{a44}
\end{equation}%
According to what happens with $x_{l}$ after update, we have two possible
evolutions:

\bigskip

B.1.1 The update of the site $l$ is such that the new value of $x_{l}$ is
larger than $\lambda $.%
\begin{equation}
x_{l}\rightarrow x_{l}^{\prime }>\lambda   \label{a45}
\end{equation}%
with%
\begin{eqnarray}
&&\textrm{probability that the secondary site }l\textrm{ belongs to the set }m%
\textrm{ and}  \nonumber \\
&&\textrm{after update }x_{l}^{\prime }\textrm{ is grater than }\lambda \textrm{ is%
}  \nonumber \\
&&\frac{m-1}{N-1}\times \left( 1-\lambda \right)   \label{a46}
\end{eqnarray}%
which reduces the number of sites in the initial set $m$ with one unit. Due
to these two updates 
\begin{eqnarray}
x_{k} &\rightarrow &x_{k}^{\prime }\ :\ m\rightarrow m\ \left( \textrm{the
main site remains under }\lambda \right)   \label{a47} \\
x_{l} &\rightarrow &x_{l}^{\prime }\ :\ m\rightarrow m-1\ \left( 
\begin{array}{c}
\textrm{the secondary site was in the }m\textrm{ family} \\ 
\textrm{but now escapes to }>\lambda 
\end{array}%
\right)   \nonumber
\end{eqnarray}%
the number of sites with $x^{\prime }s$ less than $\lambda $ is%
\begin{eqnarray}
&&\textrm{number of sites with }x\textrm{ less than }\lambda \textrm{ , at time }%
t+1\textrm{ is}\   \nonumber \\
&&m-1  \label{a48}
\end{eqnarray}%
The transition is the following%
\begin{equation}
P_{m}\left( t\right) \rightarrow P_{m-1}\left( t+1\right)   \label{a49}
\end{equation}%
which means a contribution to%
\begin{equation}
P_{m-1}\left( t+1\right)   \label{a50}
\end{equation}%
and the contribution is to $M_{m-1,m}$.%
\begin{eqnarray}
&&\lambda \times \frac{m-1}{N-1}\times \left( 1-\lambda \right)   \label{a51}
\\
&=&\frac{m-1}{N-1}\left( 1-\lambda \right) \lambda   \nonumber
\end{eqnarray}

\bigskip

B.1.2. The update of the site $l$ is such that the new value of $x_{l}$ is
smaller than $\lambda $.%
\begin{equation}
x_{l}\rightarrow x_{l}^{\prime }<\lambda   \label{a52}
\end{equation}%
with%
\begin{eqnarray}
&&\textrm{probability that the secondary site }l\textrm{ belongs to the set }m%
\textrm{ and}  \nonumber \\
&&\textrm{after update }x_{l}^{\prime }\textrm{ is less than }\lambda \textrm{ is}
\nonumber \\
&&\frac{m-1}{N-1}\times \lambda   \label{a53}
\end{eqnarray}%
which maintains the number of sites in the initial set $m$ . Due to these
two updates 
\begin{eqnarray}
x_{k} &\rightarrow &x_{k}^{\prime }\ :\ m\rightarrow m\ \ \left( \textrm{the
main site remains in }m\right)   \label{a54} \\
x_{l} &\rightarrow &x_{l}^{\prime }\ :\ m\rightarrow m\ \left( \textrm{the
secondary site remains within }m\right)   \nonumber
\end{eqnarray}%
the number of sites with $x^{\prime }s$ less than $\lambda $ is%
\begin{eqnarray}
&&\textrm{number of sites with }x\textrm{ less than }\lambda \textrm{ , at time }%
t+1\textrm{ is}\   \label{a55} \\
&&m  \nonumber
\end{eqnarray}%
The transition is%
\begin{equation}
P_{m}\left( t\right) \rightarrow P_{m}\left( t+1\right)   \label{a56}
\end{equation}%
which means a contribution to%
\begin{equation}
P_{m}\left( t+1\right)   \label{a57}
\end{equation}%
and the contribution is to $M_{m,m}$.%
\begin{eqnarray}
&&\lambda \times \frac{m-1}{N-1}\times \lambda   \label{a58} \\
&=&\frac{m-1}{N-1}\lambda ^{2}  \nonumber
\end{eqnarray}

\bigskip

B.2. Assume that 
\begin{equation}
\textrm{secondary, interacting, site }x_{l}>\lambda   \label{a59}
\end{equation}%
Then it is one of the $N-m$ sites that are all characterized at $t$ as
having $x_{i}>\lambda $.

The probability for this is%
\begin{equation}
\frac{N-m}{N-1}\left( \textrm{probability to choose }l\textrm{ between the }N-m%
\textrm{ sites with }x_{i}>l\textrm{ at }t\right)   \label{a60}
\end{equation}%
Then we have two possible evolutions:

\bigskip

B.2.1 The update of the site $l$ (which belonged to $N-m$) is such that the
new value of $x_{l}$ is larger than $\lambda $.%
\begin{equation}
x_{l}\rightarrow x_{l}^{\prime }>\lambda   \label{a61}
\end{equation}%
with%
\begin{eqnarray}
&&\textrm{probability that the secondary site }l\textrm{ belongs to the set }N-m%
\textrm{ and}  \nonumber \\
&&\textrm{after update }x_{l}^{\prime }\textrm{ is grater than }\lambda \textrm{ is%
}  \nonumber \\
&&\frac{N-m}{N-1}\times \left( 1-\lambda \right)   \label{a62}
\end{eqnarray}%
which keeps constant the number of sites in the initial set $m$. Due to
these two updates 
\begin{eqnarray}
x_{k} &\rightarrow &x_{k}^{\prime }\ :\ m\rightarrow m  \label{a63} \\
x_{l} &\rightarrow &x_{l}^{\prime }\ :\ m\rightarrow m  \nonumber
\end{eqnarray}%
the number of sites with $x^{\prime }s$ less than $\lambda $ is%
\begin{eqnarray}
&&\textrm{number of sites with }x\textrm{ less than }\lambda \textrm{ , at time }%
t+1\textrm{ is}\   \nonumber \\
&&m  \label{a64}
\end{eqnarray}%
The transition is%
\begin{equation}
P_{m}\left( t\right) \rightarrow P_{m}\left( t+1\right)   \label{a65}
\end{equation}%
which means a contribution to%
\begin{equation}
P_{m}\left( t+1\right)   \label{a66}
\end{equation}%
and the contribution is to $M_{m,m}$.%
\begin{eqnarray}
&&\lambda \times \frac{N-m}{N-1}\times \left( 1-\lambda \right)   \label{a67}
\\
&=&\frac{N-m}{N-1}\left( 1-\lambda \right) \lambda   \nonumber
\end{eqnarray}

\bigskip

B.2.2. As assumed in the class B.2 the secondary, interacting site has the
barrier $>\lambda $, which means that it does not belong to the family $m$
like $x_{k}$, but to $N-m$. The update of the site $l$ (which belonged to $%
N-m$) is such that the new value of $x_{l}$ is smaller than $\lambda $.%
\begin{equation}
x_{l}\rightarrow x_{l}^{\prime }<\lambda   \label{a68}
\end{equation}%
with%
\begin{eqnarray}
&&\textrm{probability that the secondary site }l\textrm{ belongs to the set }N-m%
\textrm{ and}  \nonumber \\
&&\textrm{after update }x_{l}^{\prime }\textrm{ is less than }\lambda \textrm{ is}
\nonumber \\
&&\frac{N-m}{N-1}\times \lambda   \label{a69}
\end{eqnarray}%
which increases the number of sites in the initial set $m$ with one unit.
Due to these two updates 
\begin{eqnarray}
x_{k} &\rightarrow &x_{k}^{\prime }\ :\ m\rightarrow m\ \ \left( \textrm{the
main site stays inside }m\textrm{ family}\right)   \label{a70} \\
x_{l} &\rightarrow &x_{l}^{\prime }\ :\ m\rightarrow m+1\ \ \left( \textrm{the
secondary site, initially out, comes to the }m\textrm{ family}\right)  
\nonumber
\end{eqnarray}%
the number of sites with $x^{\prime }s$ less than $\lambda $ is%
\begin{eqnarray}
&&\textrm{number of sites with }x\textrm{ less than }\lambda \textrm{ , at time }%
t+1\textrm{ is}\   \nonumber \\
&&m+1  \label{a71}
\end{eqnarray}%
The transition is%
\begin{equation}
P_{m}\left( t\right) \rightarrow P_{m+1}\left( t+1\right)   \label{a72}
\end{equation}%
which means a contribution to%
\begin{equation}
P_{m+1}\left( t+1\right)   \label{a73}
\end{equation}%
and the contribution is to $M_{m+1,m}$.%
\begin{eqnarray}
&&\lambda \times \frac{N-m}{N-1}\times \lambda   \label{a74} \\
&=&\frac{N-m}{N-1}\lambda ^{2}  \nonumber
\end{eqnarray}

\bigskip

\bigskip

\subsection{Results}

Now let us collect the results, \emph{i.e.} write the expression of the
elements of the matrix of transition $M_{n,m}$. They connect the state $m$
at time $t$ (whose probability is $P_{m}\left( t\right) $) with the state $n$
at time $t+1$, whose probability is $P_{n}\left( t+1\right) $. We have seen
that the transitions having $n$ as final state can only originate from
states of the small set $n$, $n\pm 1$, $n-2$. Then we can replace the
generic notation $m$ with the appropriate value from this set. The Table
below lists the connections that are possible as transitions.

\bigskip

\begin{tabular}{|l|}
\hline
$A.\ x_{k}^{\prime }>\lambda \ $%
\begin{tabular}{l}
$A1.\ x_{l}<\lambda \ 
\begin{tabular}{l}
$A1.1$ \\ 
$A1.2$%
\end{tabular}%
\ 
\begin{tabular}{l}
$x_{l}^{\prime }>\lambda $ \\ 
$x_{l}^{\prime }<\lambda $%
\end{tabular}%
\ 
\begin{tabular}{l}
$\frac{n-1}{N-1}\left( 1-\lambda \right) ^{2}$ \\ 
$\frac{n-1}{N-1}\lambda \left( 1-\lambda \right) $%
\end{tabular}%
\ 
\begin{tabular}{l}
$n-2$ \\ 
$n-1$%
\end{tabular}%
$ \\ 
$A2.\ x_{l}>\lambda \ 
\begin{tabular}{l}
$A2.1$ \\ 
$A2.2$%
\end{tabular}%
\ 
\begin{tabular}{l}
$x_{l}^{\prime }>\lambda $ \\ 
$x_{l}^{\prime }<\lambda $%
\end{tabular}%
\ 
\begin{tabular}{l}
$\frac{N-n}{N-1}\left( 1-\lambda \right) ^{2}$ \\ 
$\frac{N-n}{N-1}\lambda \left( 1-\lambda \right) $%
\end{tabular}%
\ 
\begin{tabular}{l}
$n-1$ \\ 
$n$%
\end{tabular}%
$%
\end{tabular}
\\ \hline
$B.\ x_{k}^{\prime }<\lambda \ $%
\begin{tabular}{l}
$B1.\ x_{l}<\lambda \ 
\begin{tabular}{l}
$B1.1$ \\ 
$B1.2$%
\end{tabular}%
\ 
\begin{tabular}{l}
$x_{l}^{\prime }>\lambda $ \\ 
$x_{l}^{\prime }<\lambda $%
\end{tabular}%
\ 
\begin{tabular}{l}
$\frac{n-1}{N-1}\lambda \left( 1-\lambda \right) $ \\ 
$\frac{n-1}{N-1}\lambda ^{2}$%
\end{tabular}%
\ 
\begin{tabular}{l}
$n-1$ \\ 
$n$%
\end{tabular}%
$ \\ 
$B2.\ x_{l}>\lambda \ 
\begin{tabular}{l}
$B2.1$ \\ 
$B2.2$%
\end{tabular}%
\ 
\begin{tabular}{l}
$x_{l}^{\prime }>\lambda $ \\ 
$x_{l}^{\prime }<\lambda $%
\end{tabular}%
\ 
\begin{tabular}{l}
$\frac{N-n}{N-1}\lambda \left( 1-\lambda \right) $ \\ 
$\frac{N-n}{N-1}\lambda ^{2}$%
\end{tabular}%
\ 
\begin{tabular}{l}
$n$ \\ 
$n+1$%
\end{tabular}%
$%
\end{tabular}
\\ \hline
\end{tabular}

\bigskip

\subsection{Calculation of the elements of the transition matrix}

Now we use this Table to produce the transition probabilities $M_{n,m}$,
with $m$ from the small set. With the destination state $n$, we add the
contributions that originate in one or several states of the set.

\subsubsection{The case $M_{n-2,n}$}

We take the single case of this type%
\begin{equation}
P_{n}\left( t\right) \rightarrow P_{n-2}\left( t+1\right)   \label{a75}
\end{equation}%
which is made by%
\[
A.1.1
\]%
with the probability%
\begin{equation}
M_{n-2,n}=\frac{n-1}{N-1}\left( 1-\lambda \right) ^{2}  \label{a76}
\end{equation}

\bigskip

\subsubsection{The case $M_{n,n}$}

We have to put together the cases where

\begin{itemize}
\item we start from $n$ sites with less than $\lambda $, and

\item we arrive at $n$ sites less than $\lambda $
\end{itemize}

The transition is%
\begin{equation}
P_{n}\left( t\right) \rightarrow P_{n}\left( t+1\right)   \label{a77}
\end{equation}

The sum is%
\begin{eqnarray}
A.2.2\ \ \  &\rightarrow &\ \ \frac{N-n}{N-1}\lambda \left( 1-\lambda
\right)   \label{a78} \\
B.1.2\ \ \  &\rightarrow &~~\frac{n-1}{N-1}\lambda ^{2}  \nonumber \\
B.2.1\ \ \  &\rightarrow &\ \ \ \frac{N-n}{N-1}\left( 1-\lambda \right)
\lambda   \nonumber
\end{eqnarray}%
\begin{eqnarray}
&&\frac{N-1-n+1}{N-1}2\lambda \left( 1-\lambda \right) +\frac{n-1}{N-1}%
\lambda ^{2}  \label{a79} \\
&=&2\lambda \left( 1-\lambda \right) +\frac{1}{N-1}\left[ -\left( n-1\right)
2\lambda \left( 1-\lambda \right) +\left( n-1\right) \lambda ^{2}\right]  
\nonumber \\
&=&2\lambda \left( 1-\lambda \right) +\frac{1}{N-1}\left( n-1\right) \left[
-2\lambda +2\lambda ^{2}+\lambda ^{2}\right]   \nonumber \\
&=&2\lambda \left( 1-\lambda \right) +\frac{1}{N-1}\left( n-1\right) \left(
3\lambda ^{2}-2\lambda \right)   \nonumber
\end{eqnarray}%
This is $P_{n,n}$ and is%
\begin{equation}
M_{n,n}=2\lambda \left( 1-\lambda \right) +\frac{1}{N-1}\left( n-1\right)
\left( 3\lambda ^{2}-2\lambda \right)   \label{a80}
\end{equation}%
and is OK.

\bigskip

\subsubsection{The case $M_{n-1,n}$}

We have to put together the cases where

\begin{itemize}
\item we start from $n$ sites with less than $\lambda $, and

\item we arrive at $n-1$ sites less than $\lambda $
\end{itemize}

The transition is%
\begin{equation}
P_{n}\left( t\right) \rightarrow P_{n-1}\left( t+1\right)   \label{a81}
\end{equation}

The sum is%
\begin{eqnarray}
A.1.2\ \ \  &\rightarrow &\ \ \frac{n-1}{N-1}\lambda \left( 1-\lambda
\right)   \label{a82} \\
A.2.1\ \ \  &\rightarrow &~~\frac{N-n}{N-1}\left( 1-\lambda \right) ^{2} 
\nonumber \\
B.1.1\ \ \  &\rightarrow &\ \ \ \frac{n-1}{N-1}\lambda \left( 1-\lambda
\right)   \nonumber
\end{eqnarray}%
\begin{eqnarray}
&&\frac{N-n}{N-1}\left( 1-\lambda \right) ^{2}+2\frac{n-1}{N-1}\lambda
\left( 1-\lambda \right)   \label{a83} \\
&=&\frac{N-1-n+1}{N-1}\left( 1-\lambda \right) ^{2}+2\frac{n-1}{N-1}\lambda
\left( 1-\lambda \right)   \nonumber \\
&=&\left( 1-\lambda \right) ^{2}+\frac{n-1}{N-1}\left[ -\left( 1-\lambda
\right) ^{2}+2\lambda \left( 1-\lambda \right) \right]   \nonumber \\
&=&\left( 1-\lambda \right) ^{2}+\frac{n-1}{N-1}\left( -1+2\lambda -\lambda
^{2}+2\lambda -2\lambda ^{2}\right)   \nonumber \\
&=&\left( 1-\lambda \right) ^{2}+\frac{n-1}{N-1}\left( -1+4\lambda -3\lambda
^{2}\right)   \nonumber
\end{eqnarray}%
This is $M_{n-1,n}$ and is%
\begin{equation}
M_{n-1,n}=\left( 1-\lambda \right) ^{2}+\frac{n-1}{N-1}\left( -1+4\lambda
-3\lambda ^{2}\right)   \label{a84}
\end{equation}%
and is OK.

\bigskip

\subsubsection{The case $M_{n+1,n}$}

We find the situation%
\[
B.2.2
\]%
with the probability 
\begin{eqnarray}
M_{n+1,n} &=&\frac{N-n}{N-1}\lambda ^{2}  \label{a85} \\
&=&\frac{N-1-n+1}{N-1}\lambda ^{2}=\lambda ^{2}-\lambda ^{2}\frac{n-1}{N-1} 
\nonumber
\end{eqnarray}%
and is OK.

\bigskip

These are cases for%
\begin{equation}
n\geq 1  \label{a86}
\end{equation}

\subsubsection{The particular case $M_{0,0}$}

This corresponds to the following situation: at time $t$ there is no site
under $\lambda $. At time $t+1$ the number of sites under $\lambda $ is not
modified, it is zero. This means that the update of $x_{k}$, the minimum
site, takes it from $>\lambda $ and keeps it somewhere $>\lambda $. For the
secondary, interacting, site, it was initially $>\lambda $ and after update
it remains $>\lambda $. The probability of this transition is the product of
two probabilities

\begin{itemize}
\item the probability that $x_{k}$ takes after update a value that is
greater than $\lambda $,%
\begin{equation}
1-\lambda   \label{a87}
\end{equation}

\item the probability that $x_{l}$ takes after update a value that is
greater than $\lambda $,%
\begin{equation}
1-\lambda   \label{a88}
\end{equation}
\end{itemize}

The product is the element of matrix%
\begin{equation}
M_{0,0}=\left( 1-\lambda \right) ^{2}  \label{a89}
\end{equation}

\subsubsection{The particular case $M_{1,0}$}

This can be obtained in two ways:

\begin{itemize}
\item the lowest site $x_{k}$ , which, - since we start from $m=0$, has at
time $t$ the value $>\lambda $, is updated to the same region, with
probability $\left( 1-\lambda \right) $. In the same time the secondary site
that initially at time $t$ is $>\lambda $ (since $m=0$) after update at time 
$t+1$ takes a value $<\lambda $, with probability $\lambda $. Then the
contribution of this situation to the matrix element $M_{1,0}$ is%
\begin{equation}
M_{1,0}^{\left( 1\right) }=\lambda \left( 1-\lambda \right)   \label{a90}
\end{equation}

\item the lowest site $x_{k}$ , which, - since we start from $m=0$, has at
time $t$ the value $>\lambda $ , is updated to the region that is under $%
\lambda $, with a probability $\lambda $. The secondary site $x_{l}$ is
initially (at $t$) $>\lambda $ and is updated remaining in this region $%
>\lambda $, with the probability $\left( 1-\lambda \right) $. The
contribution to the matrix element $M_{1,0}$ is%
\begin{equation}
M_{1,0}^{\left( 2\right) }=\lambda \left( 1-\lambda \right)   \label{a91}
\end{equation}
\end{itemize}

Finally we get the total matrix element%
\begin{equation}
M_{1,0}=2\lambda \left( 1-\lambda \right)   \label{a92}
\end{equation}

\bigskip

\subsubsection{The particular case $M_{2,0}$}

It is easy to see that%
\begin{equation}
M_{2,0}=\lambda ^{2}  \label{a93}
\end{equation}

\bigskip

This determines completely the set of transition probabilities (matrix
elements $M_{j,k}$).

\bigskip

\subsection{Equations connecting the probabilities of the states at the
update transition}

The knowledge of the elements of the transition matrix allows to write in
detail the equations connecting the probabilties.%
\begin{equation}
P_{n}\left( t+1\right) =\sum\limits_{m=0}^{N}M_{n,m}P_{m}\left( t\right) 
\label{a94}
\end{equation}%
To use the results obtained above, we start with the lowest cases%
\begin{eqnarray}
P_{0}\left( t+1\right)  &=&M_{0,0}P_{0}\left( t\right) +M_{0,1}P_{1}\left(
t\right) +M_{0,2}P_{2}\left( t\right)   \label{a95} \\
P_{1}\left( t+1\right)  &=&M_{1,0}P_{0}\left( t\right) +M_{1,1}P_{1}\left(
t\right) +M_{1,2}P_{2}\left( t\right)   \nonumber \\
P_{2}\left( t+1\right)  &=&M_{2,0}P_{0}\left( t\right) +M_{2,1}P_{1}\left(
t\right) +M_{2,2}P_{2}\left( t\right) +M_{2,3}P_{3}\left( t\right)  
\nonumber
\end{eqnarray}

\bigskip

First equation%
\begin{eqnarray}
P_{0}\left( t+1\right)  &=&M_{0,0}P_{0}\left( t\right) +M_{0,1}P_{1}\left(
t\right) +M_{0,2}P_{2}\left( t\right)   \label{a96} \\
&=&\left( 1-\lambda \right) ^{2}P_{0}\left( t\right)   \nonumber \\
&&+\left[ \left( 1-\lambda \right) ^{2}+\frac{n-1}{N-1}\left( -1+4\lambda
-3\lambda ^{2}\right) \right] _{n=1}P_{1}\left( t\right)   \nonumber \\
&&+\left[ \frac{n-1}{N-1}\left( 1-\lambda \right) ^{2}\right]
_{n=1}P_{2}\left( t\right)   \nonumber
\end{eqnarray}%
and we take the limit $N\rightarrow \infty $,%
\begin{equation}
P_{0}\left( t+1\right) =\left( 1-\lambda \right) ^{2}\left[ P_{0}\left(
t\right) +P_{1}\left( t\right) \right]   \label{a97}
\end{equation}

\bigskip

The second equation%
\begin{eqnarray}
P_{1}\left( t+1\right)  &=&M_{1,0}P_{0}\left( t\right) +M_{1,1}P_{1}\left(
t\right) +M_{1,2}P_{2}\left( t\right)   \label{a98} \\
&=&2\lambda \left( 1-\lambda \right) P_{0}\left( t\right)   \nonumber \\
&&+\left[ 2\lambda \left( 1-\lambda \right) +\frac{1}{N-1}\left( n-1\right)
\left( 3\lambda ^{2}-2\lambda \right) \right] _{n=1}P_{1}\left( t\right)  
\nonumber \\
&&+\left[ \left( 1-\lambda \right) ^{2}+\frac{n-1}{N-1}\left( -1+4\lambda
-3\lambda ^{2}\right) \right] _{n=2}P_{2}\left( t\right)   \nonumber
\end{eqnarray}%
\begin{eqnarray}
P_{1}\left( t+1\right)  &=&2\lambda \left( 1-\lambda \right) \left[
P_{0}\left( t\right) +P_{1}\left( t\right) \right]   \label{a99} \\
&&+\left[ \left( 1-\lambda \right) ^{2}+\frac{1}{N-1}\left( -1+4\lambda
-3\lambda ^{2}\right) \right] P_{2}\left( t\right)   \nonumber
\end{eqnarray}%
at the limit%
\begin{equation}
N\rightarrow \infty   \label{a100}
\end{equation}%
the last term vanishes and we obtain%
\begin{equation}
P_{1}\left( t+1\right) =2\lambda \left( 1-\lambda \right) \left[ P_{0}\left(
t\right) +P_{1}\left( t\right) \right] +\left( 1-\lambda \right)
^{2}P_{2}\left( t\right)   \label{a101}
\end{equation}

\bigskip

The third equation is%
\begin{eqnarray}
P_{2}\left( t+1\right)  &=&M_{2,0}P_{0}\left( t\right) +M_{2,1}P_{1}\left(
t\right) +M_{2,2}P_{2}\left( t\right) +M_{2,3}P_{3}\left( t\right) 
\label{a102} \\
&=&\lambda ^{2}P_{0}\left( t\right)   \nonumber \\
&&+\left[ \lambda ^{2}-\lambda ^{2}\frac{n-1}{N-1}\right] _{n=1}P_{1}\left(
t\right)   \nonumber \\
&&+\left[ 2\lambda \left( 1-\lambda \right) +\frac{1}{N-1}\left( n-1\right)
\left( 3\lambda ^{2}-2\lambda \right) \right] _{n=2}P_{2}\left( t\right)  
\nonumber \\
&&+\left[ \left( 1-\lambda \right) ^{2}+\frac{n-1}{N-1}\left( -1+4\lambda
-3\lambda ^{2}\right) \right] _{n=3}P_{3}\left( t\right)   \nonumber
\end{eqnarray}%
\begin{eqnarray}
&&P_{2}\left( t+1\right)   \label{a103} \\
&=&\lambda ^{2}P_{0}\left( t\right) +\lambda ^{2}P_{1}\left( t\right)  
\nonumber \\
&&+\left[ 2\lambda \left( 1-\lambda \right) +\frac{1}{N-1}\left( 3\lambda
^{2}-2\lambda \right) \right] P_{2}\left( t\right)   \nonumber \\
&&+\left[ \left( 1-\lambda \right) ^{2}+\frac{2}{N-1}\left( -1+4\lambda
-3\lambda ^{2}\right) \right] P_{3}\left( t\right)   \nonumber
\end{eqnarray}%
When 
\begin{equation}
N\rightarrow \infty   \label{a104}
\end{equation}%
the last terms in the paranthesis vanish%
\begin{eqnarray}
&&P_{2}\left( t+1\right)   \label{a105} \\
&=&\lambda ^{2}\left[ P_{0}\left( t\right) +P_{1}\left( t\right) \right]  
\nonumber \\
&&+2\lambda \left( 1-\lambda \right) P_{2}\left( t\right)   \nonumber \\
&&+\left( 1-\lambda \right) ^{2}P_{3}\left( t\right)   \nonumber
\end{eqnarray}

\bigskip

The equation for 
\begin{equation}
n\geq 3  \label{a106}
\end{equation}

In order to write this equation we have to count the transitions that are
possible when we only have $K$ sites involved in updates. The first update
is the lowest $x_{i}$ and the second is the interacting site $x_{l}$.

To reach $n$ at $t+1$ the following possibilities exist%
\begin{eqnarray}
\textrm{start from}\ \ n-1\ \  &:&\ \ M_{n,n-1}  \label{a107} \\
\textrm{start from}\ \ n\ \  &:&\ M_{n,n}  \nonumber \\
\textrm{start from}\ \ n+1\  &:&\ M_{n,n+1}  \nonumber
\end{eqnarray}%
There is another possibility, to start from $n+2$ and consider the update of
the main site $x_{k}$ that removes it from $<\lambda $ and places it above;
and update of the interacting site $x_{l}$ takes it from $<\lambda $ and
places it to $>\lambda $. This means reduction of the $n+2$ sites under $%
\lambda $ to $n$ under $\lambda $. The transition matrix is%
\begin{equation}
M_{n,n+2}  \label{a108}
\end{equation}%
which is obtained from%
\begin{equation}
M_{m-2,m}=\frac{m-1}{N-1}\left( 1-\lambda \right) ^{2}  \label{a109}
\end{equation}%
taking 
\begin{equation}
m=n+2  \label{a110}
\end{equation}%
which means%
\begin{equation}
M_{n,n+2}=\frac{n+1}{N-1}\left( 1-\lambda \right) ^{2}  \label{a111}
\end{equation}%
The limit $N\rightarrow \infty $ makes this transition element to vanish.

Then%
\begin{equation}
P_{n}\left( t+1\right) =M_{n,n-1}P_{n-1}\left( t\right) +M_{n,n}P_{n}\left(
t\right) +M_{n,n+1}P_{n+1}\left( t\right)   \label{a112}
\end{equation}%
We have to adapt the transition matrix elements to the limit $N\rightarrow
\infty $.

\bigskip

The probability of transition $M_{n,n-1}$ must be calculated from 
\begin{equation}
M_{n+1,n}=\lambda ^{2}-\lambda ^{2}\frac{n-1}{N-1}  \label{a113}
\end{equation}%
by the replacement $n\rightarrow n-1$, 
\begin{equation}
M_{n,n-1}=\lambda ^{2}-\lambda ^{2}\frac{n-2}{N-1}  \label{a114}
\end{equation}%
by taking the limit $N\rightarrow \infty $, and obtain%
\begin{equation}
M_{n,n-1}=\lambda ^{2}  \label{a115}
\end{equation}

\bigskip

The probability of transition $M_{n,n}$ becomes, after taking the limit $%
N\rightarrow \infty $, 
\begin{eqnarray}
M_{n,n} &=&2\lambda \left( 1-\lambda \right) +\frac{1}{N-1}\left( n-1\right)
\left( 3\lambda ^{2}-2\lambda \right)   \label{a116} \\
&\rightarrow &2\lambda \left( 1-\lambda \right)   \nonumber
\end{eqnarray}

\bigskip

The probability of transition $M_{n,n+1}$ is first obtained:%
\begin{eqnarray}
M_{n-1,n} &=&\left( 1-\lambda \right) ^{2}+\frac{n-1}{N-1}\left( -1+4\lambda
-3\lambda ^{2}\right)   \label{a117} \\
\textrm{after replecement }n &\rightarrow &n+1\ \ \textrm{becomes}  \nonumber \\
M_{n,n+1} &=&\left( 1-\lambda \right) ^{2}+\frac{n}{N-1}\left( -1+4\lambda
-3\lambda ^{2}\right)   \nonumber
\end{eqnarray}%
and note that at the limit $N\rightarrow \infty $ it becomes%
\begin{equation}
M_{n-1,n}\sim \left( 1-\lambda \right) ^{2}  \label{a118}
\end{equation}

\bigskip

We introduce these results in the expression for $P_{n}\left( t+1\right) $, 
\begin{eqnarray}
P_{n}\left( t+1\right)  &=&\lambda ^{2}P_{n-1}\left( t\right)   \label{a119}
\\
&&+2\lambda \left( 1-\lambda \right) P_{n}\left( t\right)   \nonumber \\
&&+\left( 1-\lambda \right) ^{2}P_{n+1}\left( t\right)   \nonumber
\end{eqnarray}%
This is the result.

\bigskip

\subsection{Avalanches in the Bak Sneppen model}

The avalanches of the Bak-Sneppen model are defined, for example, in \citet*%
{paczuskibakmaslov}. \citet*{deboer1} defines a $\lambda $-avalanche as:
\textquotedblleft\emph{an evolution taking place between two successive
times where the number }$n$ \emph{of sites lower than }$\lambda $ \underline{%
\emph{vanishes}}\textquotedblright.

To make practical this definition one considers that an avalanche has
started $t$ temporal steps ago. This can be considered time $0$.

One defines the probability $Q_{n}\left( t\right) $ of having $n$ sites with
barriers $x_{i}<\lambda $, conditioned by the situation that at $t$ there
was \emph{no} site less than $\lambda $.

The probability $Q_{n}\left( t\right) $ verifies the same equation like $%
P_{n}\left( t\right) $ but with the constraint%
\begin{equation}
M_{0,n}\rightarrow 0  \label{a120}
\end{equation}%
This means that the transition (matrix element) from the state with $n$
sites $<\lambda $ to the state with $0$ sites less than $\lambda $ is \emph{%
zero.} This condition eliminates the possibility that from the state with a
non-zero number of sites less than $\lambda $ the system cannot evolve to
the state with $0$ such sites, since if this were possible the avalanche
would be terminated.

We examine in what conditions the avalanche terminates at time $t$.

Consider the probability $Q_{n}\left( t-1\right) $ of having $n$ sites less
than $\lambda $ at time $t-1$.

The update from $t-1$ to $t$ replaces two sites: the lowest $x_{k}$ and the
interacting $x_{l}$, chosen at random. The condition that the avalanche
terminates at $t$ is that there is no more a site lower than $\lambda $, so
the transition from $x_{k}<\lambda $ at $t-1$ to $x_{k}^{\prime }$ should
move it to higher than $\lambda $ values. The probability is $\left(
1-\lambda \right) $.

Now we recognize that there may be two initial states.

\subsubsection{Case A.}

The initial state, at $t-1$ consists of only one site less than $\lambda $
and this inevitably is the lowest, $x_{k}$. After update it will move to $%
>\lambda $ with probability $\left( 1-\lambda \right) $.

However this is not all.

According to the algorithm we have to update another site, the one which is
in interaction with $x_{k}$. This site, $x_{l}$ necessarly is NOT in the
region $<\lambda $ as assumed. Then it is in the region $>\lambda $ but it
makes a transition in the same region. This is with probability $\left(
1-\lambda \right) $.

The probability of transition in this case A is%
\begin{equation}
\left( 1-\lambda \right) ^{2}  \label{a121}
\end{equation}

The initial state was%
\begin{equation}
Q_{1}\left( t-1\right)   \label{a122}
\end{equation}

\subsubsection{Case B}

The initial state at $t-1$ consists of two sites, which necessarly are

\begin{itemize}
\item the lowest, $x_{k}$ which after update goes to $>\lambda $ with
probability $\left( 1-\lambda \right) $;

\item the secondary or interacting:

\begin{itemize}
\item the probability to chose (at random) the secondary one is uniform over
the $N-1$ sites which - possibly - interact with $k$.

\item after update will move to $>\lambda $, with probability $\left(
1-\lambda \right) $.
\end{itemize}
\end{itemize}

It results that in the case B we have the result%
\begin{equation}
\frac{1}{N-1}\left( 1-\lambda \right)   \label{a123}
\end{equation}

The initial state was described by the function%
\begin{equation}
Q_{2}\left( t-1\right)   \label{a124}
\end{equation}

\bigskip

There is no other case. Since we canot assume the existence of more than two
sites $<\lambda $ since we have chosen that only two sites are updated : the
lowest one and the secondary, interacting.

\bigskip

Summing over the two cases%
\begin{eqnarray}
q\left( t\right)  &=&\left( 1-\lambda \right) ^{2}Q_{1}\left( t-1\right) 
\label{a125} \\
&&+\frac{\left( 1-\lambda \right) ^{2}}{N-1}Q_{2}\left( t-1\right)  
\nonumber
\end{eqnarray}

\bigskip

The numerical simulation of this algorithm has led to the result%
\begin{equation}
q\left( t\right) \sim \frac{1}{t^{3/2}}  \label{a126}
\end{equation}%
and is also derived analytically by \ \citet*{paczuskimaslovbak}, \citet*%
{deboer1}, \citet*{deboer2}.

\bigskip

\subsection{The limit of a large number of sites}

This means to take%
\begin{equation}
N\rightarrow \infty   \label{a127}
\end{equation}

For this we re-examine the equations for $P_{n}\left( t\right) $.

\subsubsection{Calculation of the probabilities $P_{n}\left( t\right) $ for $%
N\rightarrow \infty $ and $t\rightarrow \infty $. Connection with random
walk with reflection}

In the work of \citet*{deboer2} it is calculated the probabilities $P_{n}$ for
the large time limit. The reference model is a Random Walk with reflection
at $n=0$.

For the convergence of the geometric sum it is assumed%
\begin{equation}
\lambda <\frac{1}{2}  \label{a128}
\end{equation}%
Then the results are%
\begin{eqnarray}
P_{0} &=&1-2\lambda   \label{a129} \\
P_{1} &=&\left( 1-2\lambda \right) \left[ \frac{1}{\left( 1-\lambda \right)
^{2}}+1\right]   \nonumber \\
P_{n} &=&\left( 1+2\lambda \right) \frac{\lambda ^{2n-2}}{\left( 1-\lambda
\right) ^{2n}}  \nonumber
\end{eqnarray}

They note that as%
\begin{equation}
\lambda \rightarrow \frac{1}{2}  \label{a130}
\end{equation}%
all probabilities vanish and this means that $n$ cannot remain finite. The
probability that $n$ is finite and not very close to zero is vanishingly
small. In physical terms this means that there will be no sites which have
the fitness value under $\lambda =1/2$. This is the expression of the fact
that the domain of fitness values under $\lambda =1/2$ is now effectively
empty.

The same final conclusion is reached after calculating the probabilities $%
P_{n}\left( t\right) $ for $t\rightarrow \infty $ at $\lambda >1/2$.

\bigskip

The probability that an avalanche starts at $0$ and ends at $t$ is%
\begin{equation}
q\left( t\right)   \label{a131}
\end{equation}

This is calculated by first obtaining the probabilities (mentioned before) $%
Q_{n}\left( t\right) $ that, with an avalanche started at time $0$, there
are at moment $t$ a number of $n$ sites that are still under $\lambda $. 
\begin{eqnarray}
Q_{1}\left( 1\right)  &=&2\lambda \left( 1-\lambda \right)   \label{a132} \\
Q_{2}\left( 1\right)  &=&\lambda ^{2}  \nonumber \\
Q_{n}\left( 1\right)  &=&0\ \ \textrm{for}\ \ n\geq 3  \nonumber
\end{eqnarray}

The result is%
\begin{equation}
Q_{n}\left( t\right) =\frac{2n\left( 2t+1\right) !}{\left( t+n+1\right)
!\left( t-n+1\right) !}\lambda ^{t+n-1}\left( 1-\lambda \right) ^{t-n+1}
\label{a133}
\end{equation}

\bigskip

Now it is possible to calculate the probability that an avalanche has
duration $t$:%
\begin{equation}
q\left( t\right) =\frac{\left( 2t\right) !}{\left( t+1\right) !t!}\lambda
^{t-1}\left( 1-\lambda \right) ^{t+1}  \label{a134}
\end{equation}

The average duration of an avalanche is%
\begin{equation}
\left\langle t\right\rangle =\sum\limits_{t=1}^{\infty }tq\left( t\right) =%
\frac{1}{1-2\lambda }  \label{a135}
\end{equation}%
and we see that it diverges for $\lambda \rightarrow 1/2$.

For large $t$ the probability of an avalanche of duration $t$, $q\left(
t\right) $ has the asymptotic form%
\begin{equation}
q\left( t\right) \sim \frac{\left( 1-\lambda \right) \left[ 4\lambda \left(
1-\lambda \right) \right] ^{t}}{\lambda \sqrt{\pi }}\frac{1}{t^{3/2}}
\label{a136}
\end{equation}%
with the limit at $\lambda \rightarrow 1/2$ given by%
\begin{equation}
q\left( t\right) \sim \frac{1}{\tau ^{3/2}}  \label{a137}
\end{equation}%
which is taken as the basis for the comparison with the statistics of the
observations.

\bigskip
\bibliographystyle{copernicus}
\bibliography{bibsocnodoi}

\newcommand{\noop}[1]{}
\begin{thebibliography}{25}
\providecommand{\natexlab}[1]{#1}
\providecommand{\url}[1]{{\tt #1}}
\providecommand{\urlprefix}{URL }
\expandafter\ifx\csname urlstyle\endcsname\relax
  \providecommand{\doi}[1]{doi:\discretionary{}{}{}#1}\else
  \providecommand{\doi}{doi:\discretionary{}{}{}\begingroup
  \urlstyle{rm}\Url}\fi

\bibitem[{Arakawa and Schubert(1974)}]{arakawa1974interaction}
Arakawa, A. and Schubert, W.~H.: Interaction of a cumulus cloud ensemble with
  the large-scale environment, Part I, Journal of the Atmospheric Sciences, 31,
  674--701, 1974.

\bibitem[{Bak(1996)}]{bak1996nature}
Bak, P.: How nature works: the science of self-organized criticality, vol. 212,
  Copernicus New York, 1996.

\bibitem[{Bengtsson et~al.(2013)Bengtsson, Steinheimer, Bechtold, and
  Geleyn}]{BechtoldCellular}
Bengtsson, L., Steinheimer, M., Bechtold, P., and Geleyn, J.: A stochastic
  parameterization for deep convection using cellular automata, Q. J. R.
  Meteorol. Soc., 139, 1533--1543, 2013.

\bibitem[{Boettcher and Paczuski(1996)}]{boettcherpaczuski}
Boettcher, S. and Paczuski, M.: Exact results for spatio-temporal correlations
  in a self-organized critical model of punctuated equilibrium, Phys. Rev.
  Lett., 76, 348--351, 1996.

\bibitem[{Cruz(1973)}]{cruz1977}
Cruz, L.: Venezuelean rainstorms as seen by radar, J. Appl. Meteorol., 12,
  119--126, 1973.

\bibitem[{de~Boer et~al.(1994)de~Boer, Derrida, Flyvbjerg, Jackson, and
  Wettig}]{deboer2}
de~Boer, J., Derrida, B., Flyvbjerg, H., Jackson, A.~D., and Wettig, T.: Simple
  model of self-organized biological evolution, Phy. Rev. Lett., 73, 906--909,
  1994.

\bibitem[{de~Boer et~al.(1995)de~Boer, Jackson, and Wettig}]{deboer1}
de~Boer, J., Jackson, A.~D., and Wettig, T.: Criticality in simple models of
  evolution, Phy. Rev. E, 51, 1059--1074, 1995.

\bibitem[{Emanuel(1987)}]{Emanuel1987}
Emanuel, K.~A.: An air–sea interaction model of intraseasonal oscillation in
  the tropics, J. Atmos. Sci., 44, 2324--2340, 1987.

\bibitem[{Ito(1995)}]{ito}
Ito, K.: Punctuated-equilibrium model of biological evolution is also a
  self-organized-criticality model of earthquakes, Phys. Rev. E, 52,
  3232--3233, 1995.

\bibitem[{Jensen(1998)}]{jensen1998self}
Jensen, H.~J.: Self-organized criticality: emergent complex behavior in
  physical and biological systems, vol.~10, Cambridge university press, 1998.

\bibitem[{Leary and {Houze Jr}(1979)}]{LearyHouze}
Leary, C. and {Houze Jr}, R.: Structure and evolution of convection in a
  tropical cloud cluster, J. Atmos. Sci., 36, 437--457, 1979.

\bibitem[{Lopez(1977)}]{Lopez1977}
Lopez, R.: The lognormal distribution and cumulus cloud populations, Mon. Wea.
  Rev., 105, 865--872, 1977.

\bibitem[{Mapes(1993)}]{MapesGregarious}
Mapes, B.: Gregarious tropical convection, J. Atmos. Sci., 50, 2026--2037,
  1993.

\bibitem[{Meinhardt(1982)}]{Meinhardt}
Meinhardt, H.: Models of biological pattern formation, Academic Press, London,
  1982.

\bibitem[{Neelin et~al.(1987)Neelin, Held, and Cook}]{NeelinHeldCook}
Neelin, J., Held, I., and Cook, K.: Evaporation-wind feedback and low-frequency
  variability in the tropical atmosphere, J. Atmos. Sci., 44, 2341--2348, 1987.

\bibitem[{Paczuski et~al.(1995)Paczuski, Bak, and Maslov}]{paczuskibakmaslov}
Paczuski, M., Bak, P., and Maslov, S.: Laws for stationary states in systems
  with extremal dynamics, Phys. Rev. Lett., 74, 4253--4256, 1995.

\bibitem[{Paczuski et~al.(1996)Paczuski, Maslov, and Bak}]{paczuskimaslovbak}
Paczuski, M., Maslov, S., and Bak, P.: Avalanche dynamics in evolution, growth
  and depinning models, Phys. Rev. E, 53, 414--443, 1996.

\bibitem[{Peters and Neelin(2006)}]{petersneelin}
Peters, O. and Neelin, J.~D.: Critical phenomena in atmospheric precipitation,
  arXiv.org cond-mat, 0606076, 1--5, 2006.

\bibitem[{Peters et~al.(2001)Peters, Hertlein, and Christensen}]{OlePeteRain}
Peters, O., Hertlein, C., and Christensen, K.: A Complexity View of Rainfall,
  Phys. Rev. Lett., 88, 018\,701, 2001.

\bibitem[{Peters et~al.(2002)Peters, Hertlein, and
  Christensen}]{peters2002complexity}
Peters, O., Hertlein, C., and Christensen, K.: A Complexity View of Rainfall,
  Physical Review Letters, 88, 18\,701, 2002.

\bibitem[{Peters et~al.(2009)Peters, Neelin, and Nesbitt}]{petersneelinnesbitt}
Peters, O., Neelin, J.~D., and Nesbitt, S.: Mesoscale convective systems and
  critical clusters, J. Atmos. Sci., 66, 2913--2924, 2009.

\bibitem[{Plant and Craig(2008)}]{plantcraig}
Plant, R. and Craig, C.: A stochastic parameterization for deep convection
  based on equilibrium statistics, J. Atmos. Sci., 65, 87--105, 2008.

\bibitem[{Sornette(2006)}]{sornette}
Sornette, D.: Critical phenomena in natural sciences, Springer series in
  synergetics, Springer, 2006.

\bibitem[{Su et~al.(2000)Su, Bretherton, and Chen}]{SuBrethertonChen}
Su, H., Bretherton, C., and Chen, S.: Self-aggregation and large-scale control
  of tropical deep convection: a modeling study, J. Atmos. Sci., 57,
  1798--1816, 2000.

\bibitem[{Yano(\noop{2013}submitted)}]{YanoDAOsubmitted}
Yano, J.: Formulation structure of the mass-flux convection parameterization,
  Dyn. Atmos.Oceans, \noop{2013}submitted.

\end{thebibliography}

\bigskip

\end{document}